\documentclass[usenatbib]{mn2e}

\usepackage{graphicx,subfigure}

\begin{document}

\title[Synthetic observations of a spiral galaxy]{Synthetic H{\sc{i}} observations of a simulated spiral galaxy} 
\author[David M. Acreman, Kevin A. Douglas, Clare L. Dobbs,
Christopher M. Brunt]
{David M. Acreman$^1$\thanks{ E-mail acreman@astro.ex.ac.uk}, Kevin
  A. Douglas$^1$, Clare L. Dobbs$^{1,2,3}$, Christopher M. Brunt$^1$\\
$^1$ School of Physics, University of Exeter, Stocker Road, Exeter EX4
4QL.\\
$^2$ Max-Planck-Institut f\"ur extraterrestrische Physik, Giessenbachstra\ss{}e, D-85748 Garching, Germany \\
$^3$ Universitats-Sternwarte M\"unchen, Scheinerstra\ss{}e 1, D-81679 M\"unchen, Germany
}

\maketitle

\begin{abstract}

  Using the {\sc torus} radiative transfer code we produce synthetic
  observations of the 21\,cm neutral hydrogen line from an SPH
  simulation of a spiral galaxy. The SPH representation of the galaxy
  is mapped onto an AMR grid, and a ray tracing method is used to
  calculate 21\,cm line emission for lines of sight through the AMR
  grid in different velocity channels and spatial pixels. The result
  is a synthetic spectral cube which can be directly compared to real
  observations. We compare our synthetic spectral cubes to
  observations of M31 and M33 and find good agreement, whereby
  increasing velocity channels trace the main disc of the galaxy.
  The synthetic data also show kinks in the velocity across the spiral
  arms, evidence of non-circular velocities. These are still present
  even when we blur our data to a similar resolution as the
  observations, but largely absent in M31 and M33, indicating those
  galaxies do not contain significant spiral shocks. Thus the detailed
  velocity structure of our maps better represent previous
  observations of the grand design spiral M81.

\end{abstract}


\section{Introduction}

The last decade has seen huge advances in computational modelling of
galaxies. Allied with observational developments this provides
increased opportunities to study and understand the interstellar
medium (ISM). Comparing models and observations provides valuable
insight into the dynamics and evolution of the ISM, in particular
translating the physics incorporated in the simulations into
observable features.

High resolution hydrodynamical and magneto-hydrodynamical models
    can now begin to examine how the interstellar medium is swept into
    giant molecular clouds, and ultimately forms stars
    \citep{Wada99,Shetty06,Wada08,Dobbs08,Tasker09,DGCK08}.  The
    ability of supernovae to trigger molecular cloud formation can
    also be studied using numerical models.
    \citep{deAvillez_2001,dib_2006}. At the same time, high resolution
    surveys now provide data on the distribution and properties of
    molecular clouds \citep{Heyer98,Eng03,Ros07,Brunt09}.  Advances in
    observational capability also provide increased insight into the
    characteristics of the ISM in external galaxies e.g. recent
    H{\sc{i}} observations of the nearby spiral galaxies M31
    \citep{chemin_2009} and M33 \citep{Putman09} and the irregular
    galaxies Holmberg~{\sc{II}} \citep{rhode_1999} and the LMC
    \citep{kim_2007}.  Additionally, statistical properties of
    H{\sc{i}} in galaxies can be studied using larger samples obtained
    through surveys \citep{walter_2008, tamburro_2009}. Thus comparing
    simulations and observations is particularly timely;
    however in order to provide a direct comparison the output of the
    simulations needs to be processed to produce the same data
    products, such as spectral line datacubes and maps, which are
    obtained from observations.

An obvious application of such synthetic observations is to test
whether numerical models can reproduce the structure of the
interstellar medium seen in observations, thus synthetic
    observations provide an important validation test of numerical
    models. Synthetic observations may also be generated with very
    high spatial resolution compared to observed data, hence
    simulations can show what we will view in nearby galaxies with
    future facilities. Moreover, synthetic observations are derived
    from simulated objects with known properties (e.g density,
    temperature, velocity) whereas when dealing with real observations
    these properties are calculated from the observations. Within the
    Milky Way we commonly rely on radial velocities to determine
    distances, however for simulations, features can readily be
    located both in velocity and cartesian space. Hence working with
    synthetic observations makes it easier for features in the
    observations to be related to the physical processes from which
    they result.  An important effect in this paper is the impact of
    velocity structure on H{\sc{i}} observations.

Previous comparisons of velocity structure in simulated spiral
    galaxies with observed data range from simply identifying clouds
and plotting their position in velocity space \citep{Dobbs06,Baba09},
to sophisticated radiative transfer models \citep{Chak09,Naray09}
and synthetic galactic plane surveys
    \citep{gomez_2004,douglas_2010}. \citet{Dobbs06} plotted the
distribution of molecular clouds in velocity space to show that they
reproduce the distribution in the Outer Galaxy reasonably
well. \cite{Baba09} plotted the location of cold gas using
the velocities of the gas in their models and an assumed rotation
curve, and highlighted the disparity with the actual location of the
gas, illustrating the so called `finger of God' effect.  On larger
scales, \citet{Naray09} and \citet{Chak09} perform analysis on SPH
(smoothed particle hydrodynamics) simulations of interacting galaxies,
using radiative transfer codes to determine CO emission and spectral
energy distributions respectively. \cite{gomez_2004} and
\cite{douglas_2010} generate synthetic galactic plane surveys and
compare their velocity structure to real observations from the Leiden/Dwingeloo H{\sc{i}}
survey and the Canadian Galactic Plane Survey respectively.

On extragalactic scales H\,{\sc{i}} is a powerful tracer used to
analyse galactic structure, e.g. tracing spiral arms and tidal tails,
and showing holes and cavities in the ISM. Synthetic H\,{\sc{i}} maps
have been constructed previously \citep{Wada00,Dib05}, though for
irregular rather than spiral galaxies. These studies compared
synthetic H\,{\sc{i}} maps with data from Holmberg II and the LMC,
concluding that the structure in these galaxies is primarily due to a
combination of turbulence and thermal and gravitational instabilities,
with supernovae contributing to a lesser degree. In this paper,
    we produce synthetic H\,{\sc{i}} maps of a simulated grand design
    spiral galaxy \citep{DGCK08} in which the structure is
instead dominated by spiral density waves, and the gas velocities
reveal information about the spiral structure. Velocity gradients
reveal streaming motions, due to the presence of spiral density waves
(or at least stellar spiral arms which rotate at lower $\Omega(r)$
than the stars, see \citet{Dobbs09,wada_2004}) which produce spiral
shocks in the gas.

As our model galaxy is a perfectly isolated, grand design system,
    which has no effects from interactions with companions or high
    velocity clouds, we expect to see only effects due to spiral
    shocks.  \citet{Visser1980} first showed kinks in the velocities
    along the spiral arms in M81, due to non-circular motions,
    typically with a magnitude of a few 10's of km s$^{-1}$
    \citep{Adler96}. We expect to see similar effects in our synthetic
    observations and our idealised model system allows us to study
    these effects without the additional complexity introduced by
    environmental interactions.

We used an SPH code to model the galaxy (unlike the grid based
calculations by \citet{Dib05} and \citet{Wada00}) which we then
combine with {\sc torus}, a grid-based radiative transfer code.  Thus
we have the further complication of converting between SPH and a grid
code. We begin in Section~\ref{sec:method} by describing the method
used to generate synthetic observations, including the SPH to grid
conversion. The technique is used to produce spectral cubes for
galaxies with orientations like M31 and M33, which are assessed
    and compared to real observations of M31 and M33, in
    Section~\ref{sec:results}, as a means of validating the
    method. Although these galaxies do not display grand design
    structure to the extent of M81 (indeed M33 is a flocculent
    spiral), we have access to high resolution H{\sc{i}} data for
    these nearby systems.

\section{Method}
\label{sec:method}

This section describes the method used to generate the synthetic
observations, starting from the SPH simulation of the spiral galaxy
and going through to the production of a 21\,cm spectral line cube. 

\subsection{SPH simulation of the spiral galaxy}

\subsubsection{Calculations}

The calculation used in this paper was set up as described in
\citet{DGCK08}, but we also provide a description below. The
calculation constitutes one of a series investigating structure and
the formation of molecular clouds in spiral galaxies, and most
notably, includes a full thermodynamical model of the ISM, and the
conversion between atomic hydrogen and molecular hydrogen, necessary
for producing our synthetic observations. We did not however include
self gravity or magnetic fields, although these have been the subject
of previous papers \citep{Dobbs08,DP08}. 
The presence of magnetic fields is expected to inhibit the
    formation of structure in the galactic disc and reduce the strength
  of spiral shocks \citep{roberts_1970,DP08}, resulting in a smoother
distribution of gas, and spiral arms which are wider and less
dense. Conversely, self-gravity will increase density in regions which
are already overdense (enhancing the formation of structure in the
gas) and increase the likelihood that gas in dense regions will be
fully molecular.

The calculation used as the basis for our synthetic observations was
performed using smoothed particle hydrodynamics (SPH), a Lagrangian
fluids method.  We modelled a gaseous disc, whilst the stellar
component was included as an external potential. This external
potential includes a stellar disc and halo, and incorporates a 4 armed
spiral component from \citet{Cox02} with a pattern speed of $2 \times
10^{-8}$~rad~yr$^{-1}$ and a pitch angle of $15^o$. The amplitude of
the stellar spiral perturbation is $1.1 \times 10^{12}$ cm$^2$
s$^{-2}$.

We did not model the whole disc, but restricted the domain to radii
between 5 and 10 kpc, primarily to increase resolution. The gas
particles were initially distributed randomly, with a scale height of
400 pc and a temperature of 7000 K.  The particles were assigned
circular velocities according to the disc potential, with a 6 km
s$^{-1}$ velocity dispersion.

The surface density of the galaxy was 10 $M_{\odot}$ pc$^2$, including
helium, which is comparable to the average surface density at the
solar radius \citep{Wolfire03}. We used 8 million particles in
the calculation thus giving a resolution of 500 M$_{\odot}$ per
particle.

We followed the thermal evolution of the gas using a model for the
heating and cooling of atomic and molecular gas taken from
\citet{Glover07}. This model incorporated many processes,
including cooling from fine structure emission, gas-grain cooling, and
heating from photoelectric emission and cosmic ray ionisation of
atomic hydrogen.

We also included the formation of molecular hydrogen, as well as
CO. We evolve the abundance of molecular gas according to a
prescription given by \cite{Bergin04}, and provide a full description
in \cite{Dobbs08}. We assign each particle a molecular gas fraction,
and calculate the rate of formation of $\rm{H}_2$ on grains, and the rate of
destruction of $\rm{H}_2$ by photodissociation and cosmic ray ionization, to
update the molecular gas fraction after each time step. We noted in our
previous paper that determining the photodissociation rate requires an
estimate of the column density in order to take into account self
shielding. As described in \cite{Dobbs08} we adopt a constant length
of 35~pc to calculate the column density, which is the average
distance to a B0 star in the Milky Way. Given the $\rm{H}_2$ fraction for a
particular particle, we determine the H{\sc{i}} density used for this paper
from $n(H)=n-2n(H_2)$, where $n(H), n(H_2)$ and $n$ are the number
densities of H{\sc{i}}, $\rm{H}_2$ and the total number density. 

\subsubsection{Evolution of galaxy disc}

The evolution of the gas disc is described fully in \citet{DGCK08} but
we also provide a brief summary here. As the disc evolves, the spiral
pattern emerges and narrow spiral arms develop. Most of the gas cools
from the initial temperature of 7000 K and typically 70 \% of the gas
(and all the gas in the spiral arms) is $<150$ K. Clumps of cold gas
accumulate into more massive clouds as they pass through a spiral
shock. The molecular gas (which constitutes around 30 \% of the total
gas) lies predominantly in the spiral arms, and typically forms from
the atomic medium over time scales of order 10 Myr. As dense clumps
leave the spiral arms, they are sheared into short spurs. Some cold
clumps, or spurs, are able to survive between the arms, though they do
not tend to contain much molecular gas.  
    We do not have observational measurements of the proportions of
    warm and cold gas in external galaxies. The fractions in
    our simulations are not dissimilar from the solar
    neighbourhood, the only region for which there is an observational 
    indication of the fractions of cold and warm HI
    \citep{Heiles2003}. In the absence of observational evidence to
    the contrary we can only assume that other nearby large spiral
    galaxies are not vastly different in terms of the amount of gas in
    different phases.

We run the simulation for a total of 320 Myr. For the comparisons in
this paper, we take the output at a time of 250 Myr, by which point
the distribution of gas in different phases, and the amount of
molecular gas, has reached a roughly steady state.

\subsection{AMR grid construction}

We use the radiative transfer code {\sc torus} \citep{Harries00} to
generate synthetic H\,{\sc{i}} observations. {\sc torus} is a
grid-based radiative transfer code that uses Adaptive Mesh Refinement
(AMR) to provide variable spatial resolution.  {\sc torus} can perform
radiative transfer calculations, using the Monte Carlo method of
\cite{Lucy}, and can also generate spectral energy distributions,
images and spectral cubes. The code has frequently been applied to
models of stellar discs and performs well in benchmark tests, even at
high optical depths \citep{pinte_2009}.

To generate a synthetic H\,{\sc{i}} data cube we first need to convert
from the particle representation of SPH to the AMR grid representation
of {\sc torus}. The {\sc torus} AMR grid is constructed using the
octree method in which the grid initially comprises eight cells (one
octal) with two cells in each spatial dimension.  The grid is refined
by specifying a condition which determines when a cell is split into a
further eight cells.  For this calculation the grid cells were split
if the mass of the cell exceeded a given limit (mass per cell
limit). A mass per cell limit gives higher spatial resolution in
regions of high density, which is similar to the effective spatial
resolution of the SPH method. A maximum H\,{\sc{i}} mass per cell of
$2.5\times 10^{36}$~g (1260~$\rm{M}_{\odot}$) was used to split the
grid, which resulted in an AMR grid comprising 678815 octals and
4751706 unique cells. This mass per cell limit corresponds to
    approximately 9 SPH particles per grid cell (for particles
    with an average atomic hydrogen fraction) and was chosen to give
    sufficient spatial resolution to represent the important features
    of the model galaxy, within the constraints of the available
    computer memory. Should a calculation require higher spatial
    resolution, the mass per cell limit can be decreased to give
    smaller cells in the AMR grid \citep{douglas_2010}.

\begin{figure}
  \centering
  \includegraphics[scale=0.3]{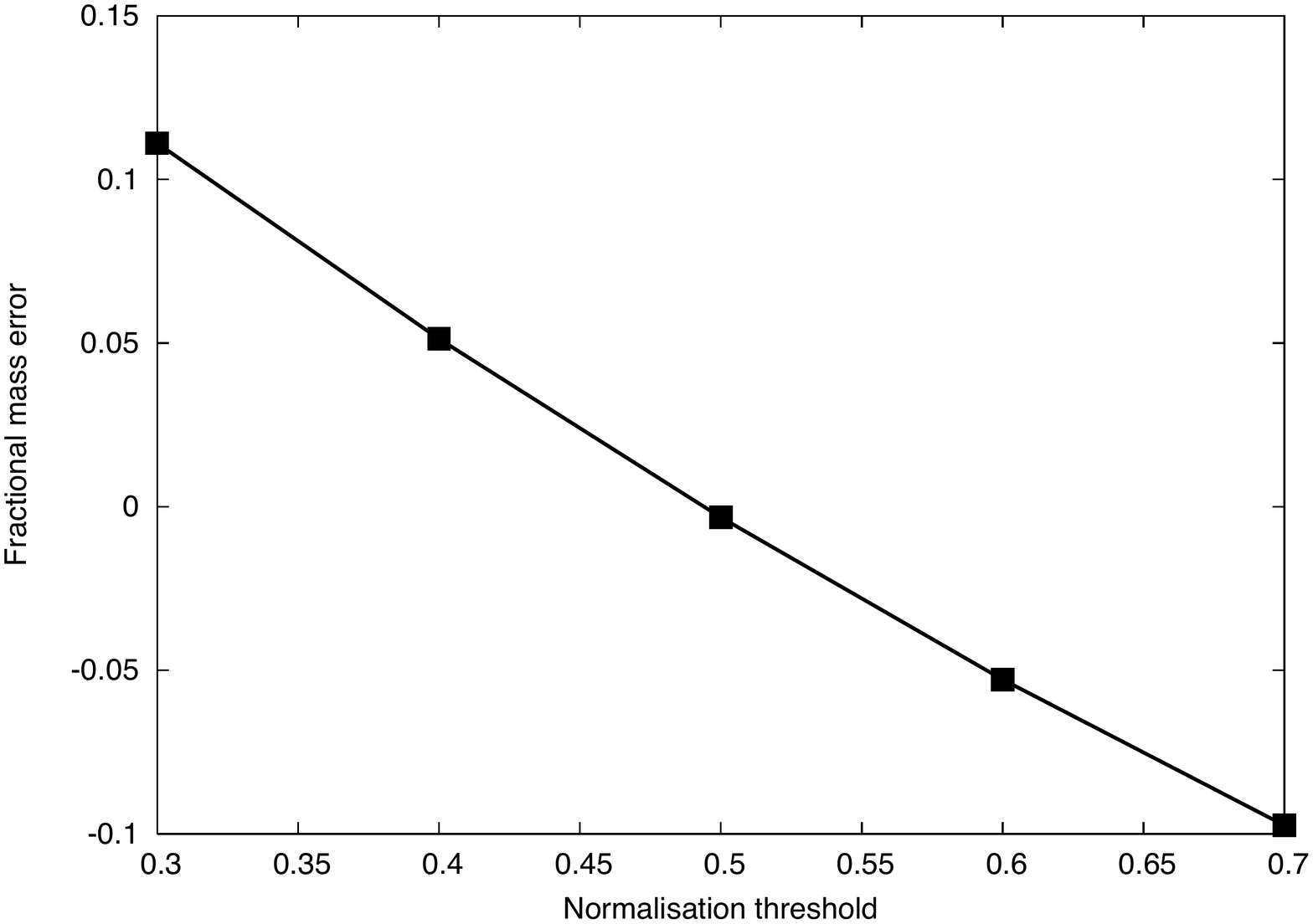}
  \caption{Fractional error in H\,{\sc{i}} mass on the AMR grid,
    relative to the H\,{\sc{i}} mass of the SPH particles, as a
    function of threshold for normalisation of the SPH kernel
    smoothing. Positive values of mass error indicate that the grid
    mass is too large. The lines are line sections joining points and
    not a best fit line.}
  \label{fig:mass_err_vs_norm}
\end{figure}

Once the grid structure has been generated H\,{\sc{i}} density,
temperature and velocity values are assigned to each cell using values
from the SPH particles, smoothed with an exponential
kernel. Construction of an AMR grid from SPH particles, using this
method, was tested by \cite{Acreman09} using an azimuthally symmetric
circumstellar disc benchmark.  In the case of a spiral galaxy it is
also necessary to represent accurately structure, such as spiral arms,
clouds and spurs, which are not present in an azimuthally symmetric
disc. We found that the method of \cite{Acreman09} gave a good
representation of structure within the disc and a good representation
of the total H\,{\sc{i}} mass provided one modification was made. The
method normalises density values by the sum of the SPH kernel weights,
if the sum of the weights exceeds a specified threshold, in order to
reduce noise in the interior of the distribution. \cite{Acreman09}
normalise when the sum of the weights exceeds 0.3, but we found this
value gave a total H\,{\sc{i}} mass which was too large, by 11 per
cent, relative to the H\,{\sc{i}} mass of the SPH particles.  The SPH
to grid conversion was performed with a range of normalisation
thresholds, in order to test the impact on the representation of the
total H\,{\sc{i}} mass, and the results are plotted in
Fig.~\ref{fig:mass_err_vs_norm}.
A normalisation threshold of 0.5 gives a mass which is too small by only
$0.3$~per~cent, so we use this value as the threshold above which the
SPH kernel smooth is normalised. 

\begin{figure*}
  \centering
  \subfigure[]{\includegraphics[scale=0.2]{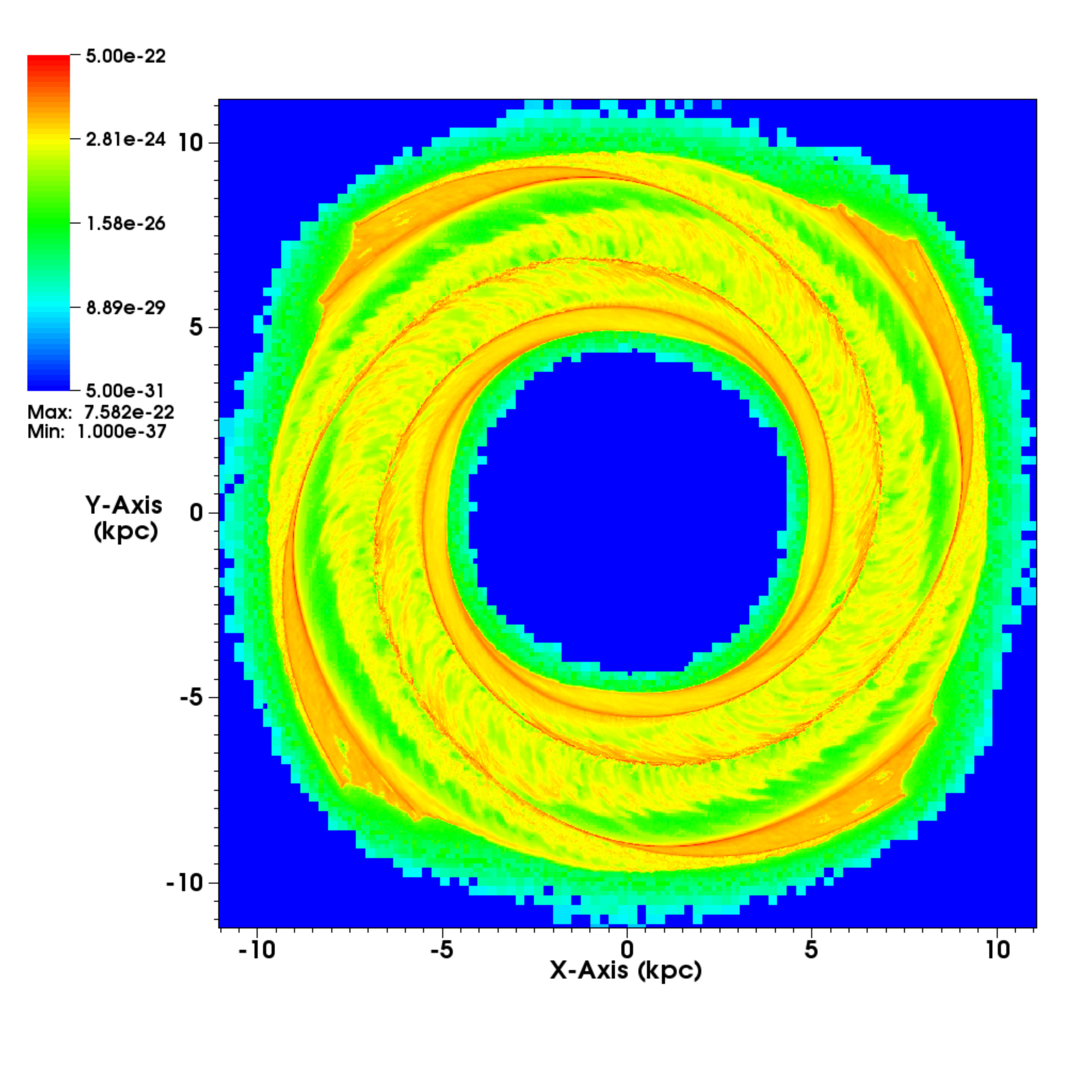}\label{fig:rho_whole_amr}}
  \subfigure[]{\includegraphics[scale=0.32]{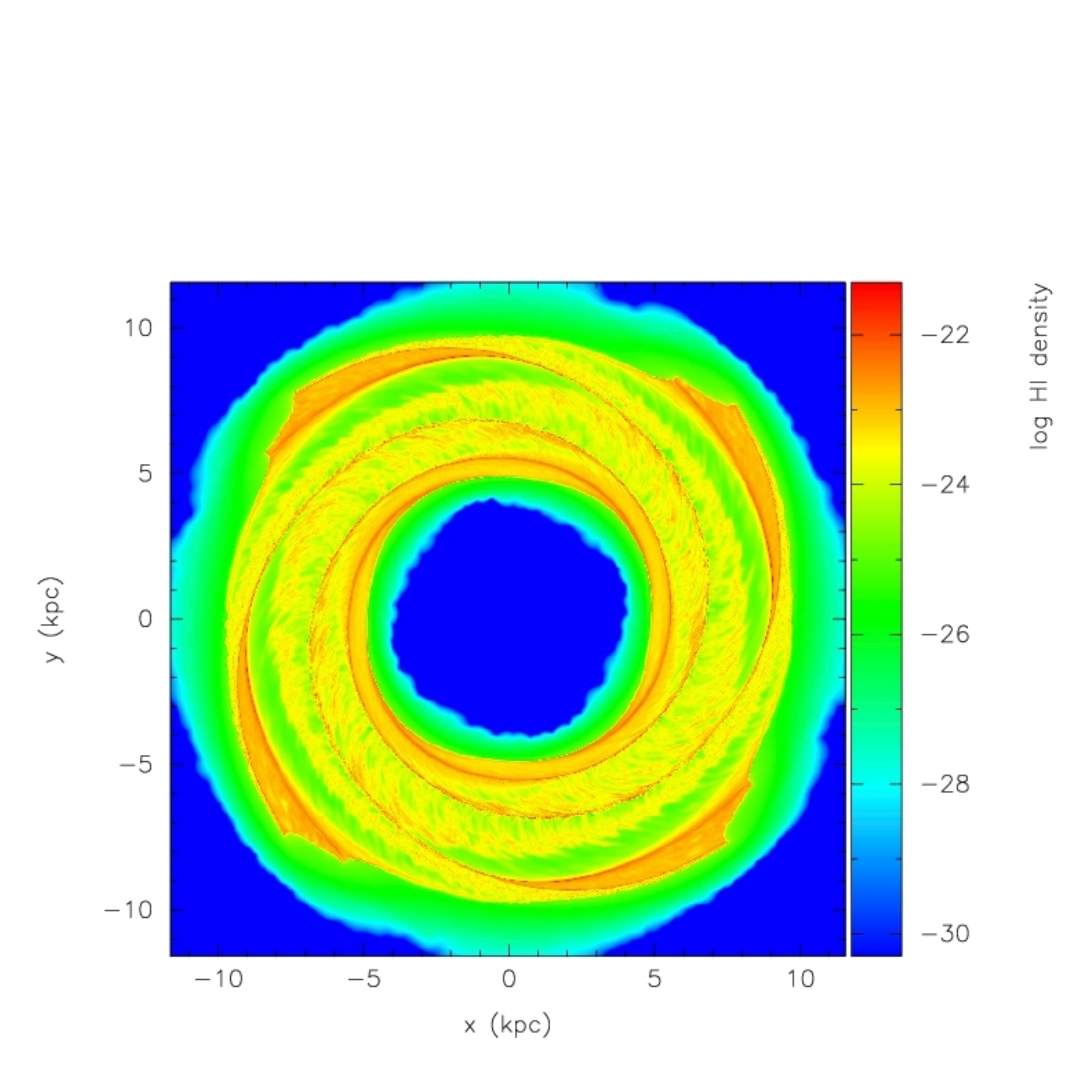}\label{fig:rho_whole_sph}}
  \caption{{H\sc{i}} density (in $\rm{g/cm}^3$) in the galaxy midplane, as represented on
    the {\sc {torus}} AMR grid (left) and as represented with SPH
    particles (right). The spiral arms are well represented in both cases
    demonstrating that the conversion from SPH particles to AMR grid
    preserves spiral arm structure.}
  \label{fig:rho_whole}
\end{figure*}

\begin{figure*}
  \centering
  \subfigure[]{\includegraphics[scale=0.2]{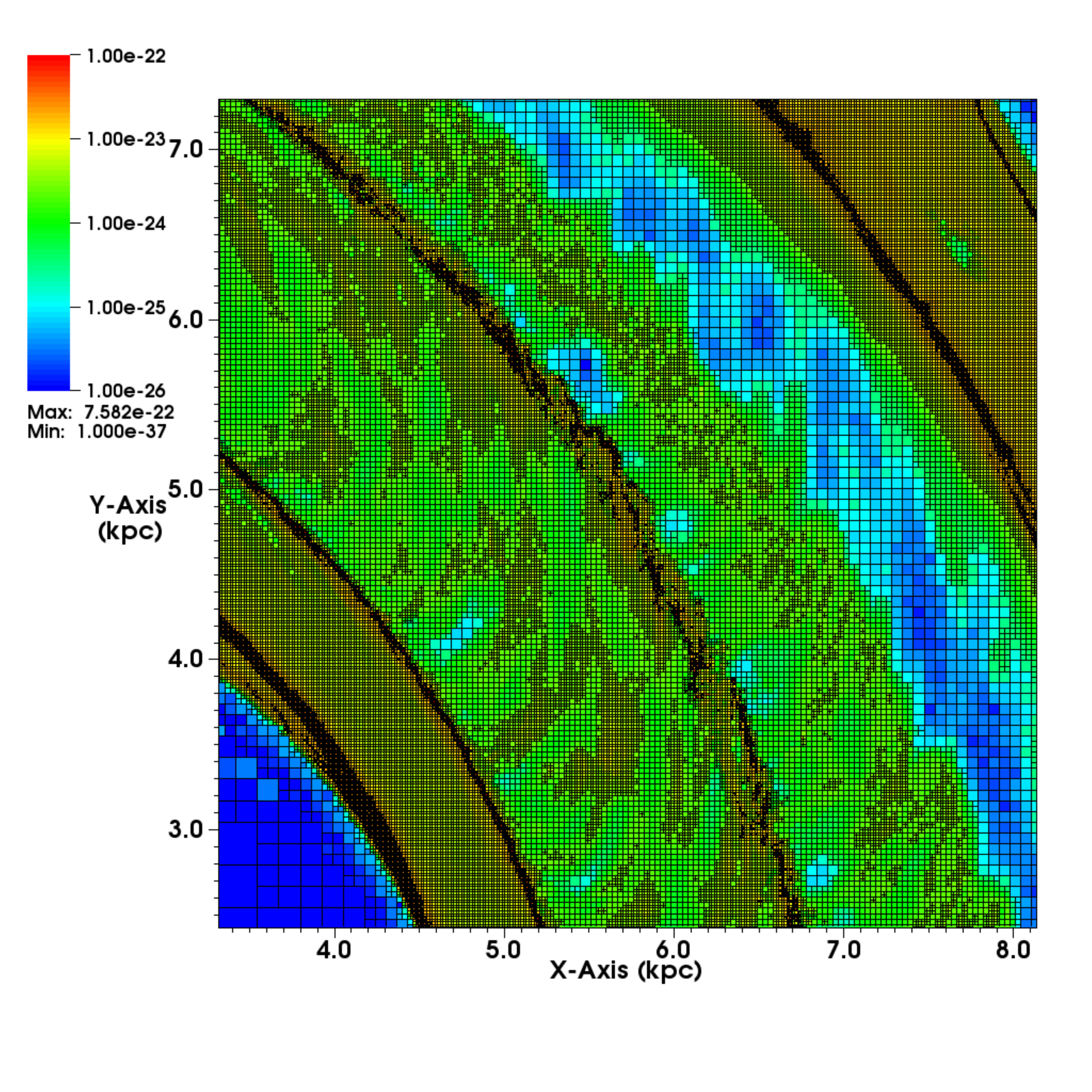}\label{fig:rho_with_mesh}}
  \subfigure[]{\includegraphics[scale=0.5]{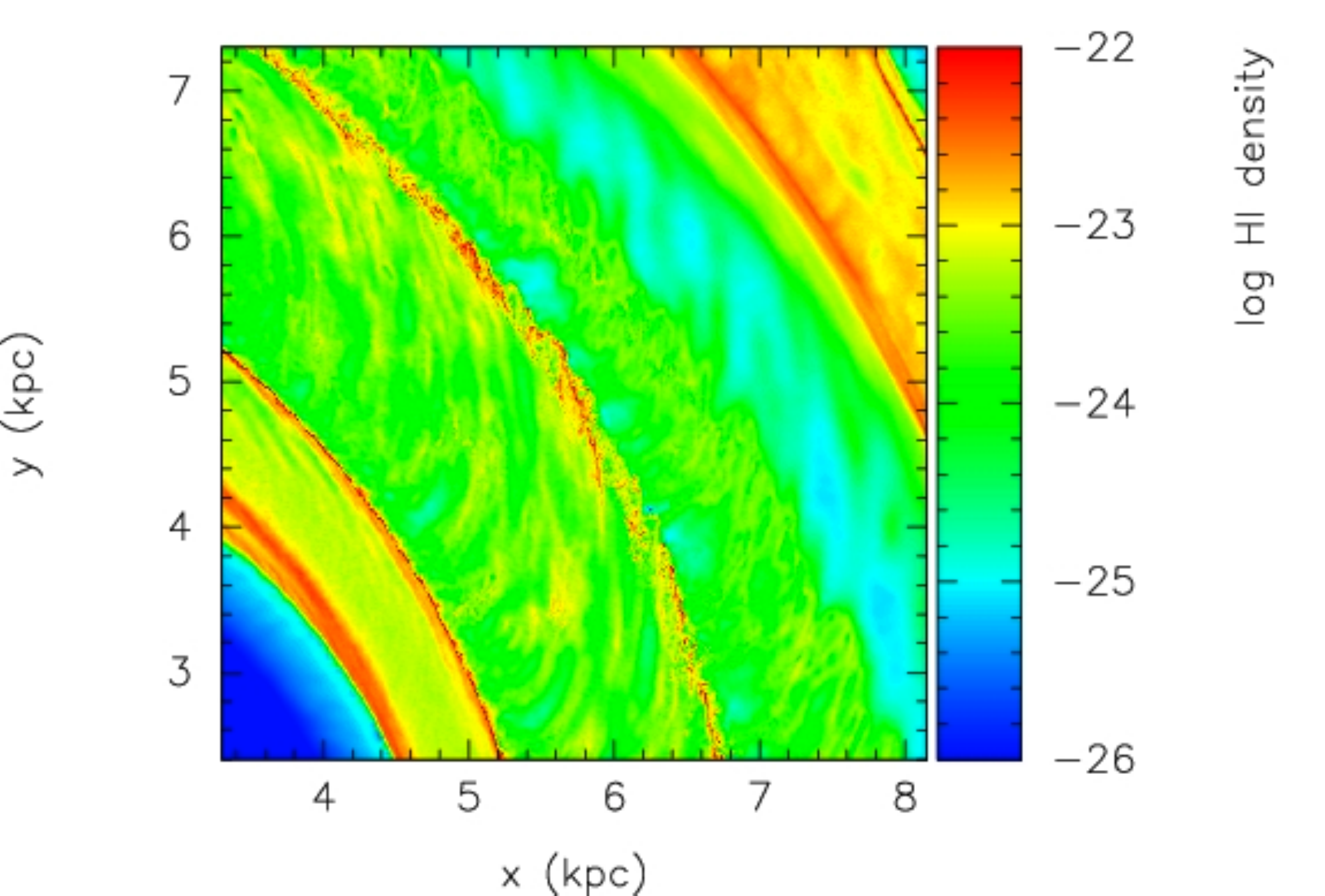}}
  \caption{
    {H\sc{i}} density (in $\rm{g/cm}^3$) in the galaxy midplane, as
    represented on the {\sc {torus}} AMR grid (left) and as
    represented with SPH particles (right). The AMR representation has
    the grid overplotted.  This figure is similar to
    Fig.~\ref{fig:rho_whole} but shows a smaller domain to emphasize
    that the smaller scale structures are well represented on the
    grid.}
  \label{fig:rho_zoom}
\end{figure*}

The H{\sc{i}} density in the galaxy midplane, as represented on the
AMR grid, is shown in Fig.~\ref{fig:rho_whole_amr}. For purposes
    of comparison a similar plot from the original SPH particles is
    shown in Fig.~\ref{fig:rho_whole_sph}, where the particles have
    been rendered using the {\sc{SPLASH}} plotting software \citep{price_2007}.
This figure shows that the spiral arms are well represented in
    both cases and the SPH particle to AMR grid conversion has been
effective in preserving these features. Figure~\ref{fig:rho_zoom}
again shows the midplane density in the grid and SPH
    representations, but for a reduced domain, with the AMR mesh over
plotted in Fig.~\ref{fig:rho_with_mesh}, to emphasize how the
enhanced AMR resolution in regions of high density enables good
resolution of small scale features.

\subsection{Opacity and emissivity calculation}

The relatively simple nature of the  H\,{\sc{i}} 21\,cm line
transition means that once the density and temperature of a cell are
known the emissivity and opacity can be calculated and stored on the
AMR grid. The opacity $\kappa_{\nu}$ is given by
\begin{equation}
\kappa_{\nu} = \frac{3 c^2 h A_0}{32 \pi k \nu_0} \frac{n(\rm{H})}{T_S} \phi
\left( v \right)
\label{eqn:opacity}
\end{equation}
\citep{rohlfs} where $c$ is the speed of light, $h$ is Planck's
constant, $A_0$ is the Einstein probability emission coefficient, $k$
is the Stefan-Boltzmann constant, and $\nu_0$ is the frequency of the
H\,{\sc{i}} transition. The number density of H\,{\sc{i}}
($n(\rm{H})$) and the excitation temperature ($T_S$) are taken from
the density and temperature values on the AMR grid. The emissivity is
calculated from the opacity using Kirchoff's law
($\epsilon_{\nu}=\kappa_{\nu} B_{\nu}$ where $B_{\nu}$ is the Planck
formula) and the Rayleigh-Jeans approximation giving
\begin{equation}
\epsilon_{\nu} = \frac{3 \nu_0 h A_0}{16 \pi
  } {n(\rm{H})} \phi
\left( v \right).
\label{eqn:emissivity}
\end{equation}

The profile function $\phi\left( v \right)$ is assumed
to be a Gaussian with a thermal line width $\sigma_{th}$ given by 
\begin{equation}
\sigma^2_{th} = \frac{k T}{m_{H}}
\end{equation}
where $T$ is the temperature of the grid cell and ${m_{H}}$ is the
mass of a hydrogen atom. Unlike \cite{Wada00} we do not add a term to
represent microturbulence as the majority of the gas in our simulation
is at temperatures where we expect the thermal line width to dominate
(see fig.~4 of \cite{DGCK08}).  As the contribution of the line
profile to the opacity and emissivity depends on the velocity channel
under consideration the line profile is calculated during the
radiative transfer step and is not stored on the AMR grid.

\subsection{Radiative transfer calculation}

Once the AMR grid is set up a spectral cube is generated by carrying
out ray tracing operations through the AMR grid.  The method is
similar to that of \cite{rundle_2009} and is summarised below.

A ray is traced through the grid, starting from the observer's
position and proceeding along the ray direction to distances further
from the observer. The contribution of the grid cell to the pixel
brightness is calculated using the emissivity and opacity taken from
the AMR grid. Each cell contributes to the intensity in the pixel but
with absorption by the accumulated optical depth between that cell and
the observer. There is an iterative procedure in which the
intensity of the pixel is updated from an old value $I_{\nu}$ to a new
value $I_{\nu} ^\prime$ with the contribution of a grid cell according to 
\begin{equation}
I_{\nu} ^\prime= I_{\nu} + S_{\nu} \left( 1 - e^{-d\tau} \right) e^ {-\tau_{\rm{total}}}
\end{equation}
where $S_{\nu} = \frac{\epsilon_{\nu}}{\kappa_{\nu}}$, $d\tau$ is the
optical depth of the cell and $\tau_{\rm{total}}$ is the total optical
depth between the cell and the observer. 

The model grid is rotated to represent a galaxy with the required
position angle and inclination angle. A number of parallel rays, one
per pixel, are traced through the grid to calculate emission for each
pixel in each velocity channel of the cube. The ray trace through each
individual cell is sub-divided into smaller steps to give better
sampling of the velocity gradient across the cell. If the velocity
difference across the cell, projected onto the direction of the ray,
is $\Delta v$ then the ray trace is decomposed into $5 \frac{\Delta
  v}{\sigma_{th}}$ separate steps.

The synthetic observations are converted to units of brightness
temperature, using the Rayleigh-Jeans approximation, by multiplying by
$\frac{\lambda^2}{2 k}$, where $\lambda$ is the wavelength of the
transition. Continuum emission from the cosmic microwave background is
included as a blackbody component with a temperature of 2.73~K. 

\section{Results}
\label{sec:results}

Our technique was used to generate synthetic spectral line data cubes
resembling radio observations of M31 and M33, two Local Group spiral
galaxies, for which arcminute-scale 21 cm observations were
available. The model galaxy was not set up to model M31 or M33
specifically, so although we expect the synthetic data to reproduce
general features seen in H{\sc{i}} observations of external galaxies
we do not make a detailed quantitative comparison with the
observations. However comparisons of the synthetic data with observed
data do provide a valuable validation test of the method.

\subsection{M31}

\begin{figure*}
  \centering
  \subfigure[Column density]{\includegraphics[scale=0.45]{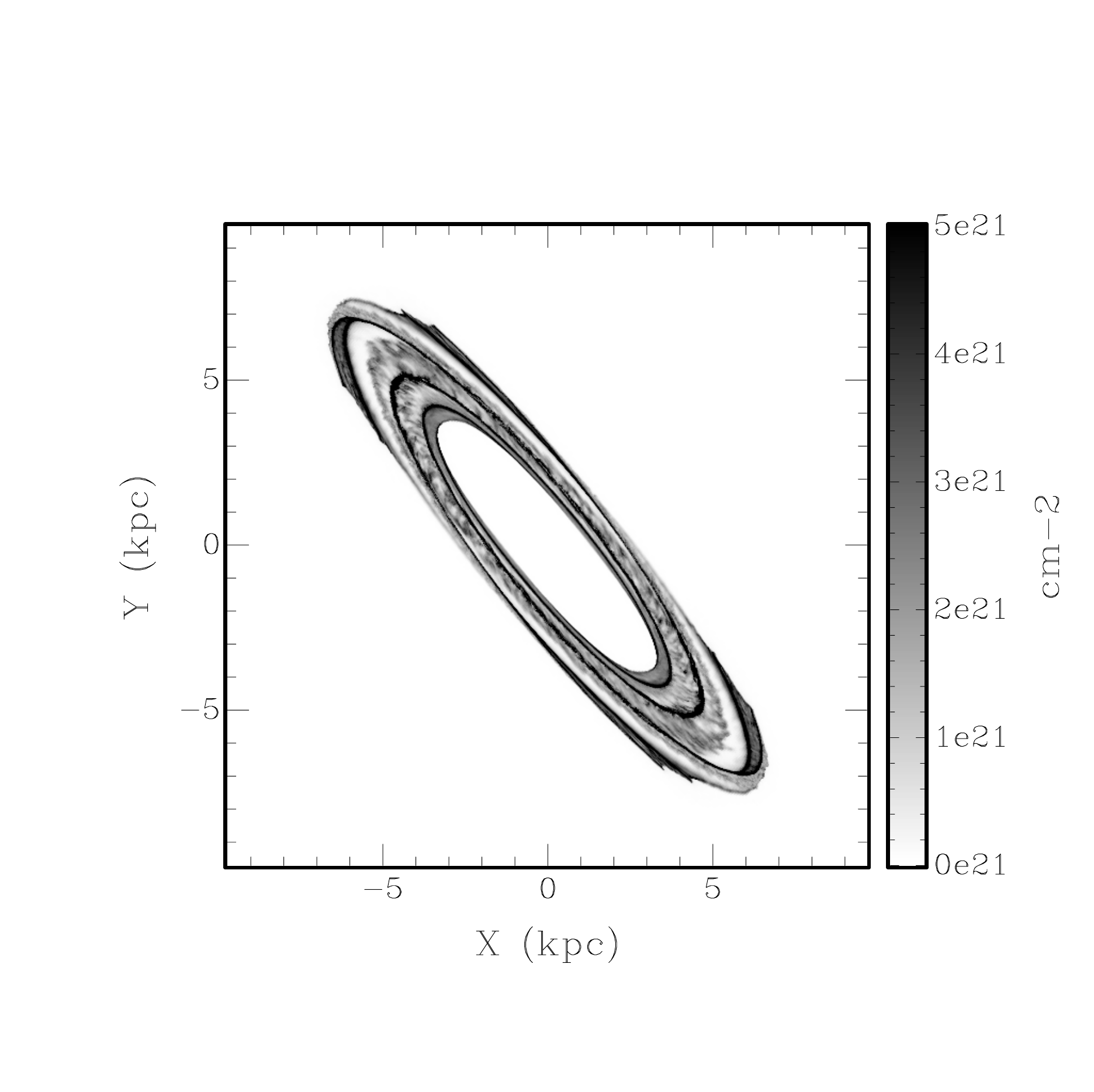}\label{fig:ncol_M31}}
  \subfigure[Simulated data blurred with 1 pixel Gaussian]{\includegraphics[scale=0.45]{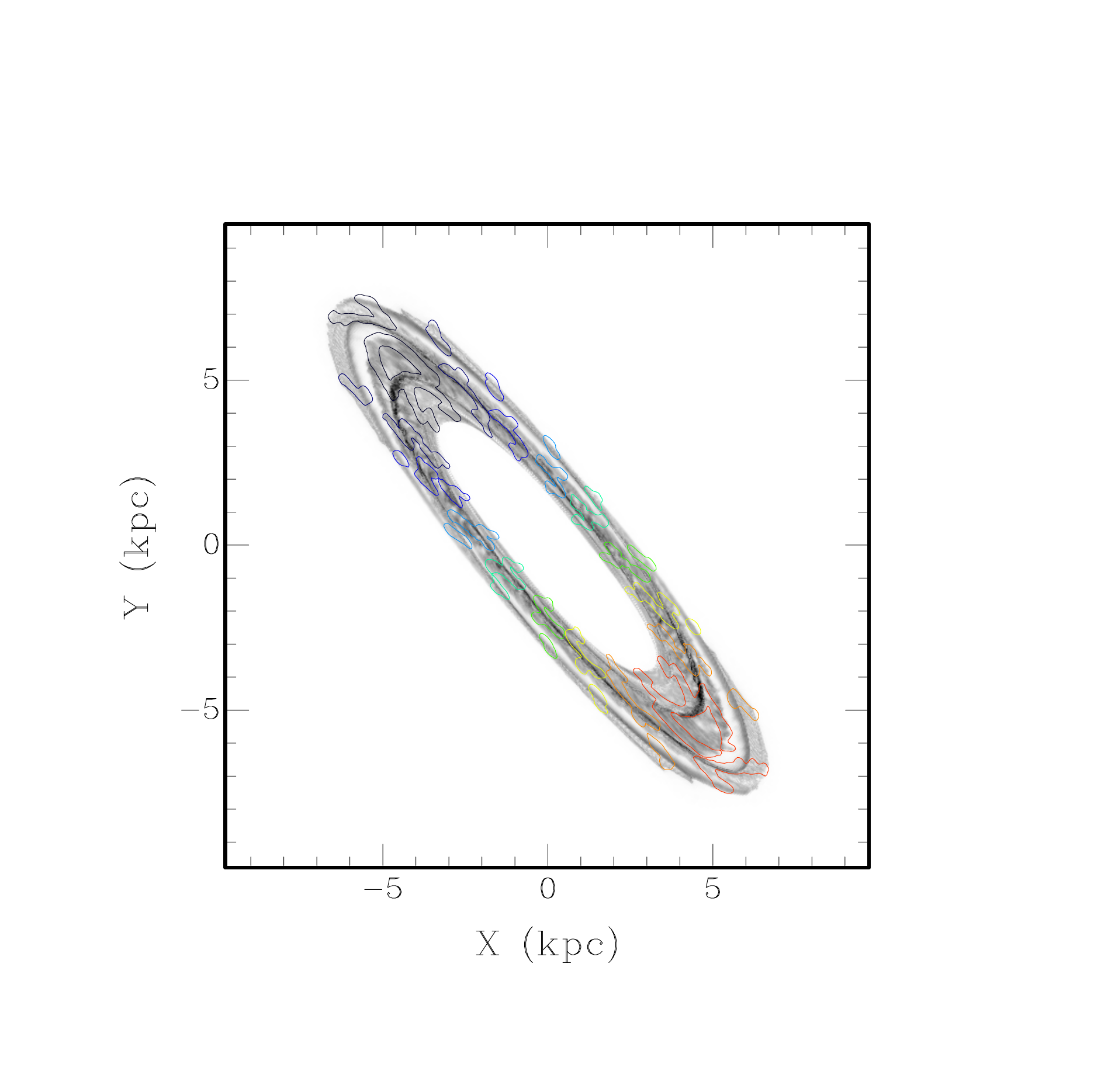}\label{fig:synth_blur1_M31}}
  \subfigure[Simulated data blurred with 3 pixel Gaussian]{\includegraphics[scale=0.45]{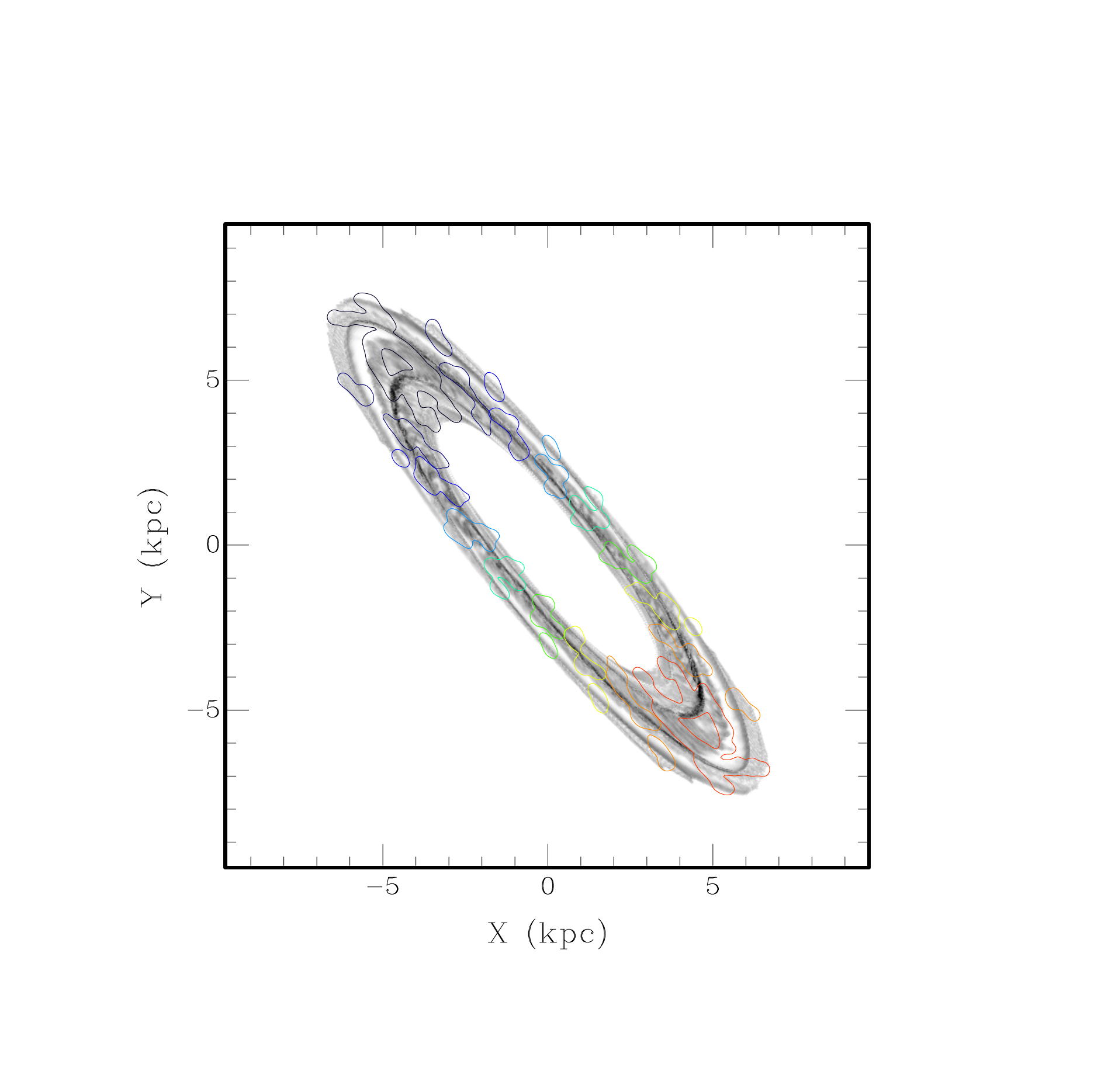}\label{fig:synth_blur3_M31}}
  \subfigure[Observed data]{\includegraphics[scale=0.425]{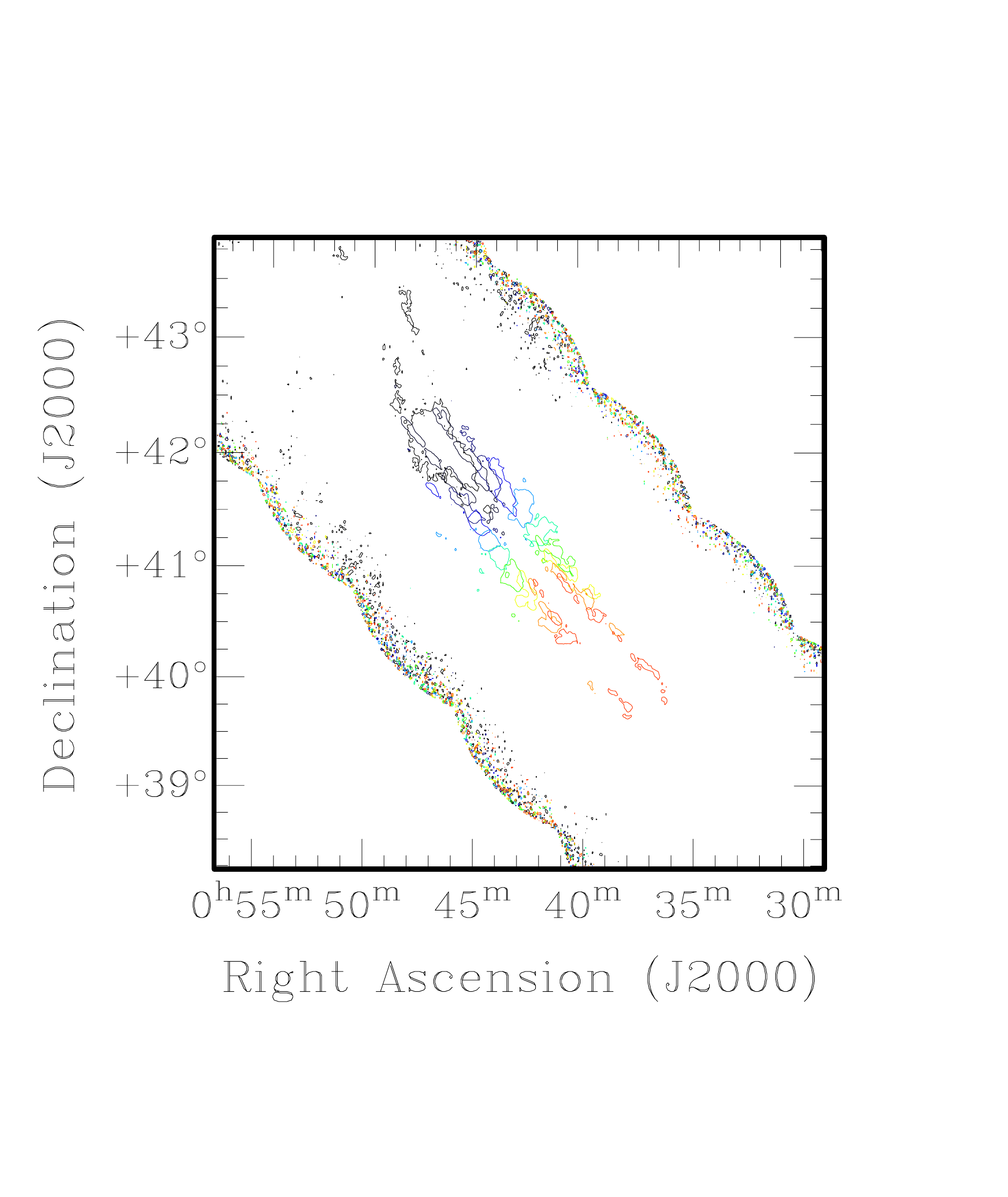}\label{fig:obs_M31}}
  \caption{Top left: column density from the model galaxy orientated
    to be like M31. Top right: contours of constant brightness
    temperature in different velocity channels (renzogram) overlaid on
    summed intensity for the model galaxy. There is one contour per
    velocity channel with different velocity channels indicated by
    different colours ($-500$ km/s for the red contour and $-100$ km/s
    for the black contour). The data have been blurred with a one pixel
    Gaussian. Lower left: as for top right but blurred with a three
    pixel Gaussian. Lower right: renzogram plot from the M31
    observation of \citealt{chemin_2009}.}
  \label{fig:M31}
\end{figure*}

The model galaxy was given an inclination angle of 77.5~degrees and a
major axis position angle of 220.0~degrees (measured clockwise from
North) in order to match the observed orientation of M31 in equatorial
co-ordinates (see figure 9 of \cite{chemin_2009}). The position
  angle of the model galaxy is offset by 180~degrees so that it
  rotates in the same direction as M31. A velocity offset of
$-300$~km/s was applied to the data cube velocity channels
consistent with the systemic velocity determined by
\cite{chemin_2009}. The orientation of the galaxy can be seen
in the column density plot in Fig.~\ref{fig:ncol_M31}. The spectral
cube was constructed using 600 spatial pixels of size
$10^{20}~\rm{cm}$ and 250 velocity channels over a range of
$840$~km/s. The velocity channels were chosen to give comparable
velocity resolution to the observations of \cite{chemin_2009}.

The emission from the synthetic data cubes is plotted as
``renzograms'' (Fig.~\ref{fig:synth_blur1_M31} and
Fig.~\ref{fig:synth_blur3_M31}) which plot a single level contour map
for several velocity channels. The contours are of constant brightness
temperature with different velocity channels shown in different
colours. The contours are at 50~km/s intervals starting at $-500$~km/s
(red contour) and ending at $-100$~km/s (black contour).  The contours
are overlaid on a grey scale plot of brightness temperature summed
over all velocity channels. These synthetic data cubes have been
spatially blurred with a Gaussian filter of 1 pixel
(Fig.~\ref{fig:synth_blur1_M31}) and 3 pixels
(Fig.~\ref{fig:synth_blur3_M31}) width to simulate the effect of
observing with an instrument with finite spatial resolution (1 pixel
corresponds to 0.14 arcmin and 3 pixels corresponds to 0.43 arcmin at
a distance of 785~kpc).
A similar plot of the observational data presented by
\cite{chemin_2009} is shown in Fig.~\ref{fig:obs_M31}. As the rotation
velocity of the model galaxy (220 km~s$^{-1}$) and the rotation
velocity of M31 (230 km/s in the outer galaxy \citealt{chemin_2009})
are similar, the velocity channels which are contoured in
Fig.~\ref{fig:obs_M31} are the same as in
Fig.~\ref{fig:synth_blur1_M31} and Fig.~\ref{fig:synth_blur3_M31}.

A good general agreement is seen between the synthetic and real data
whereby the galaxy's main outer disc is traced as the velocity
channels increase. The synthetic data with the least blurring
(i.e. the highest spatial resolution) show structure associated with
the spiral shock in the disc. This structure is not particularly
evident in the observed data for M31. This may be because of lack of
resolution (it is also more difficult to distinguish the spiral
structure in our 3 pixel blurred image) or simply because the spiral
shocks are too weak or nonexistent. In fact \cite{Efremov09} argues
that there is a spiral shock in one arm, but not the other, based on
observations of stellar gradients across spiral arms and the
distribution of star complexes, so the presence of spiral shocks in
M31 is rather ambiguous.

\begin{figure}
  \centering
\includegraphics[scale=0.3]{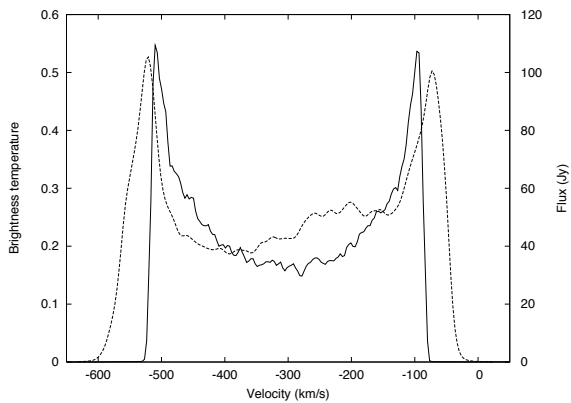}
\caption{A spatially averaged line profile from the simulated M31
  galaxy (solid line, left side y-axis). The average is taken over the
  whole data cube and background subtracted. A similar spatially
  averaged line profile from the observations of \citealt{chemin_2009}
  is also plotted (dashed line, right side y-axis). }
  \label{fig:M31_line_profile}
\end{figure}

A line profile from the synthetic data, blurred with a one pixel
Gaussian and spatially summed over the whole galaxy, is plotted in
Fig.~\ref{fig:M31_line_profile} (solid line). A similar observed
global line profile from the data of \cite{chemin_2009} is also
plotted (dashed line).
As the circular velocity of the model galaxy is similar to the
rotation velocity of M31 the modelled line profile can be
quantitatively compared to the observed line profile. Both profiles
show peaks at around $-100$~km~s$^{-1}$ and $-500$~km~s$^{-1}$
although M31 has a slightly greater circular velocity, so the peaks
    are slightly further apart. The observed profile is more extended
    than the synthetic profile, in that the emission falls away less
    rapidly towards -600~km/s and zero; the effect is more pronounced
    towards more negative velocities. Figure 7 of \cite{chemin_2009}
    shows that material with velocities around -600 km/s is to be
    found at approximately one third of the disc outer radius (0.8
    degrees from the centre of M31). This material would lie within
    the 5~kpc inner radius of our model galaxy and would not be
    represented. The observed profile also shows asymmetric structure
    between the peaks, which is not present in the synthetic line
    profile, which we also expect to come from the central regions of
    the galaxy. There is likely to be complex structure present in the
    inner regions of a real galaxy, which is not represented in our
    model galaxy, and this structure can noticeably affect the global
    line profile.

\begin{figure}
\centering
\includegraphics[scale=0.3]{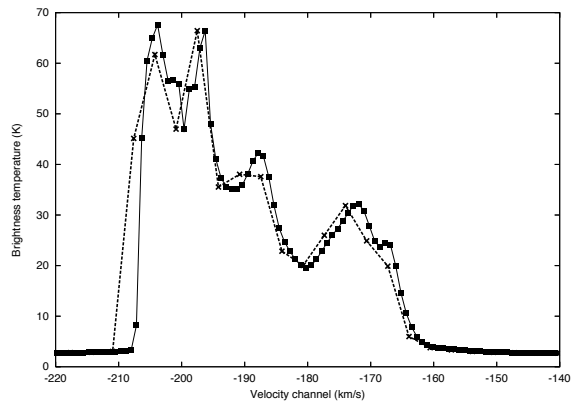}
\caption{Line profiles at a point from M31 synthetic data with 1000 velocity
  channels (solid line and squares) and 250 velocity channels (dashed
  line and crosses.)}
\label{fig:line_x250_y420}
\end{figure}

Many of the line profiles from individual pixels contain structure
which is better represented in data with higher velocity
resolution. The synthetic data cube was regenerated using 1000
velocity bins, over the same velocity range, and the effect on an
example line profile can be seen in Fig.~\ref{fig:line_x250_y420}
(solid line with squares is 1000 velocity channels, dashed line with
crosses is 250 velocity channels). This profile is taken from
$x=-1.59$~kpc, $y=3.91$~kpc which is in one of the spiral arms.
The line profile from the data cube with 250 channels shows that the
shape of the line is much less accurately represented than with 1000
channels and some features of the line (e.g between $-165$ and $-170$
km/s are not represented in the lower resolution case. The presence of
structure with intrinsically narrow velocity widths was observed in
VLA observations of M31 by \cite{braun_1990} and our model galaxy
exhibits similar narrow velocity widths in its spectral features.  For
accurate quantitative work the data which are output from the
radiative transfer code must be represented with sufficient velocity
resolution to represent accurately these narrow features of the line
profiles.

\subsection{M33}

The model galaxy was given an inclination angle of 50.0~degrees and a
major axis position angle of 20.0~degrees (measured clockwise from
North) in order to match the observed orientation of M33 in equatorial
co-ordinates. These values are representative of the {H\sc{i}}
    distribution fitted by \cite{corbelli_1997}. A velocity offset
of $-180$~km/s was applied to the data cube velocity channels in order
to match the observed  approach velocity\citep{corbelli_1997}.
The cube comprises 600 spatial pixels of size $10^{20}~\rm{cm}$ and
500 velocity channels over a range of 360~km/s.

\begin{figure*}
  \centering
  \subfigure[Column density]{\includegraphics[scale=0.45]{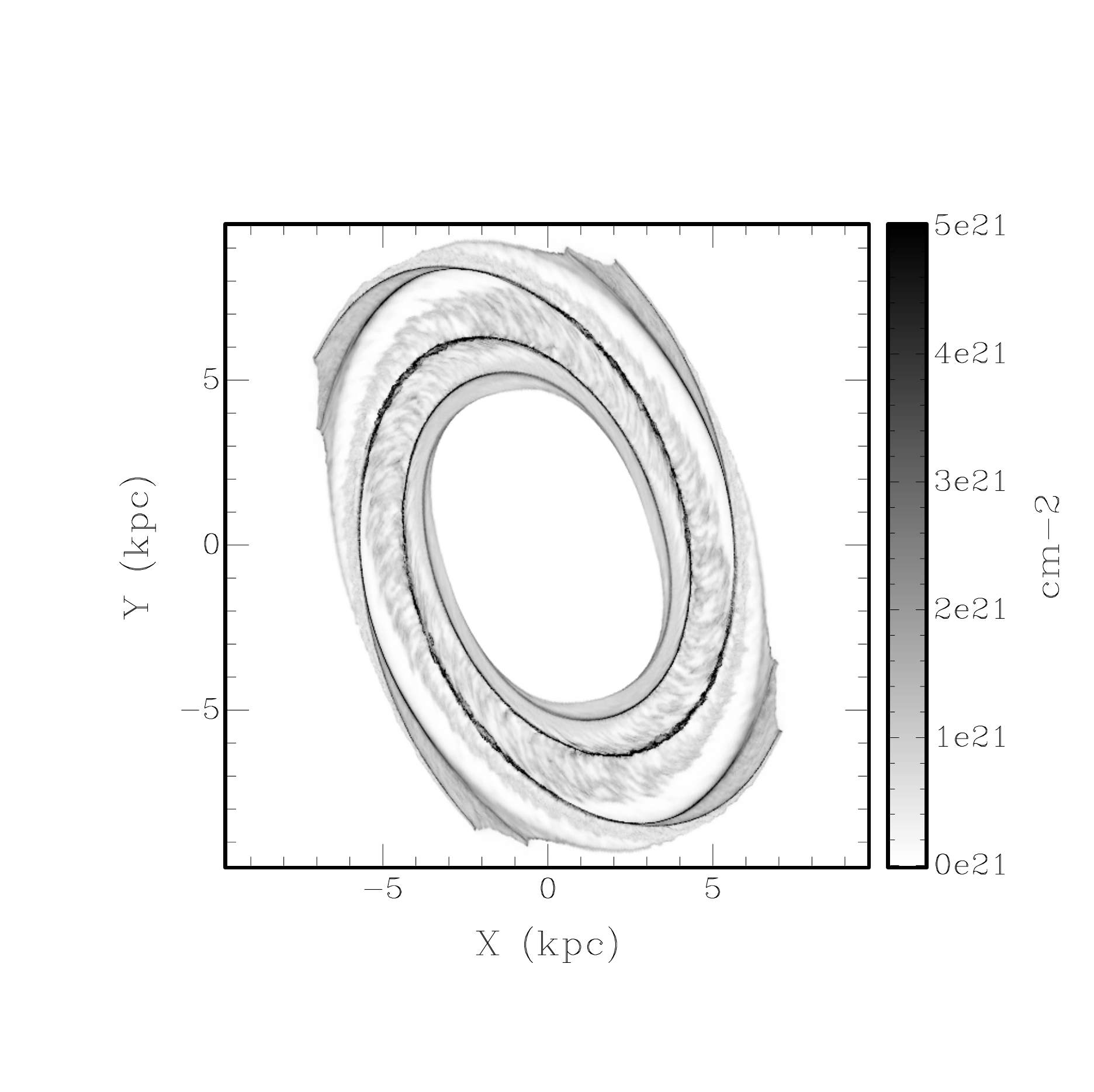}\label{fig:ncol_M33}}
  \subfigure[Simulated data blurred with 1 pixel Gaussian]{\includegraphics[scale=0.45]{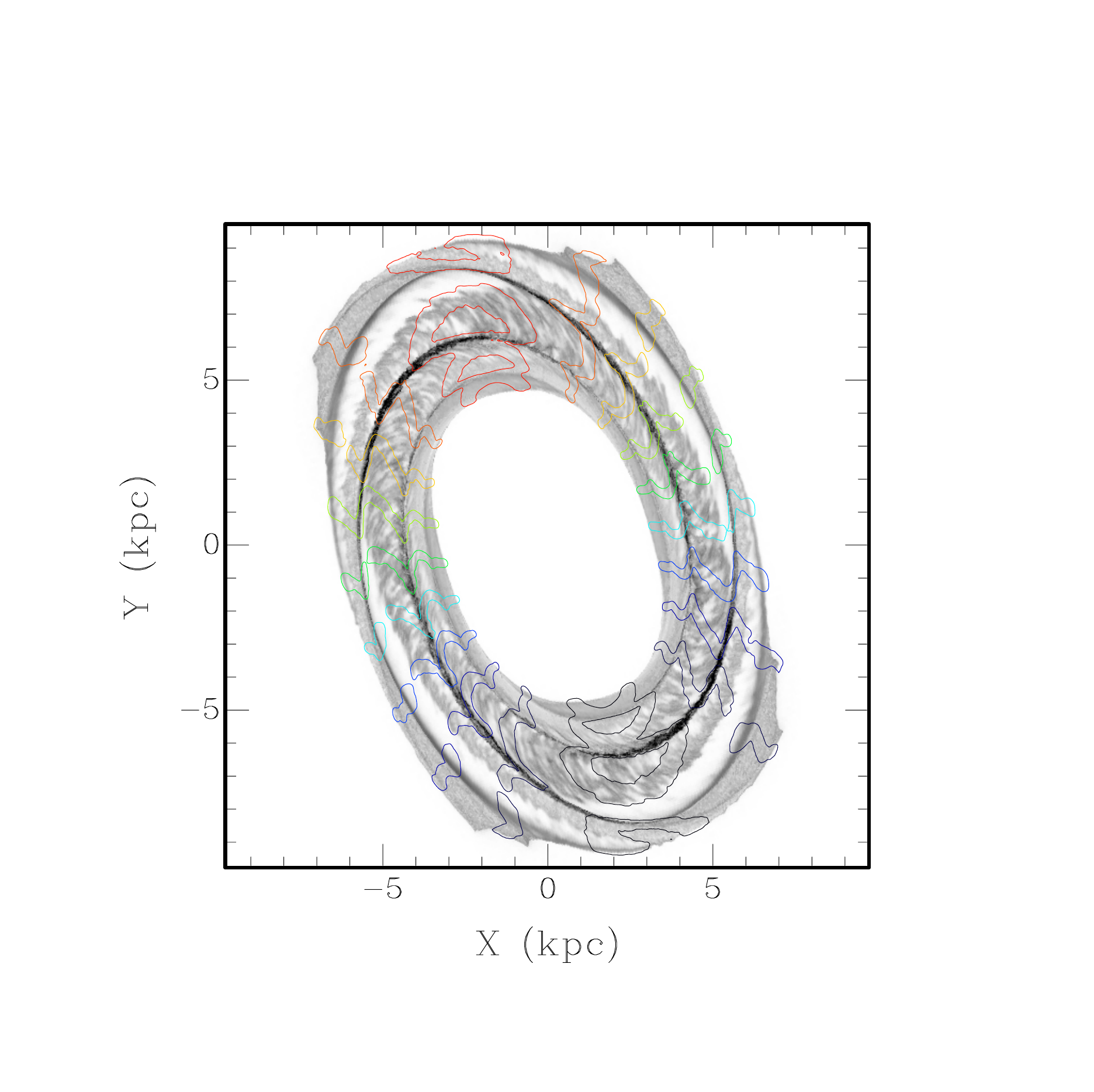}\label{fig:synth_blur1_M33}}
  \subfigure[Simulated data blurred with 6 pixel Gaussian]{\includegraphics[scale=0.45]{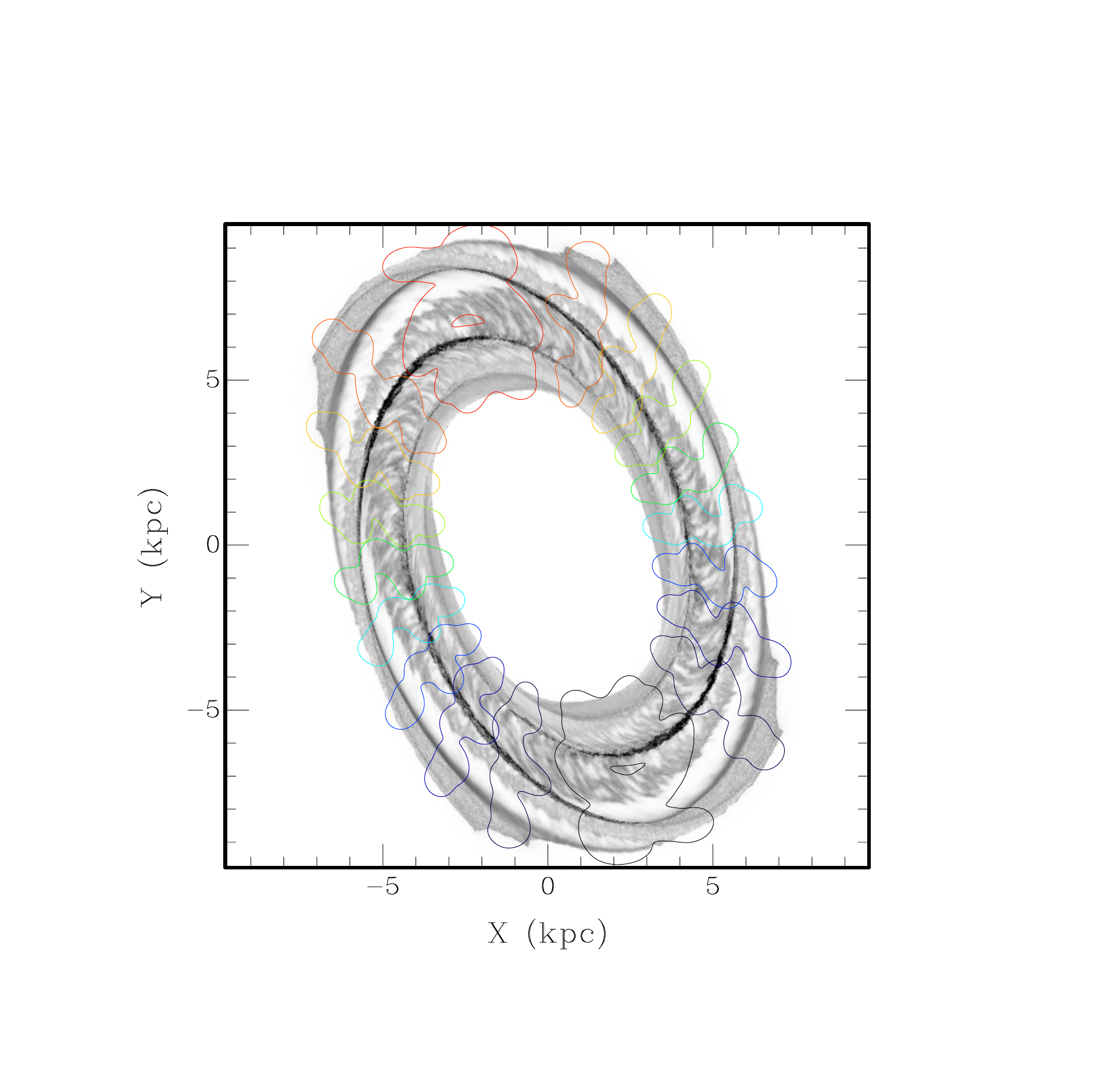}\label{fig:synth_blur6_M33}}
  \subfigure[Observed data]{\includegraphics[scale=0.35]{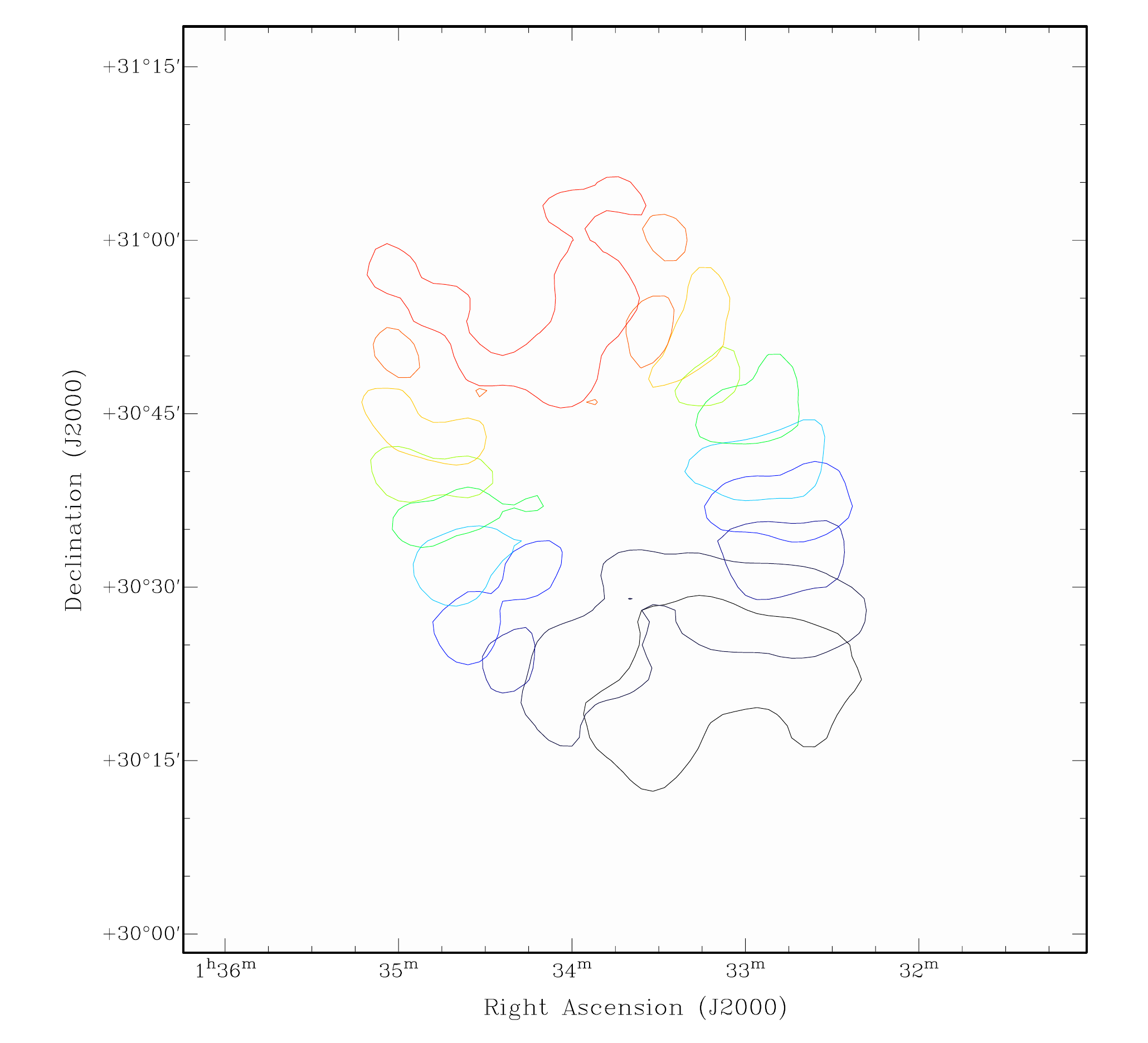}\label{fig:obs_M33}}
  \caption{Top left: column density from the model galaxy orientated
    to be like M33. Top right: contours of constant brightness
    temperature in different velocity channels (renzogram) overlaid on
    summed intensity for the model galaxy. There is one contour per
    velocity channel with different velocity channels indicated by
    different colours. The data have been blurred with a one pixel
    Gaussian. Lower left: as for top right but blurred with a six
    pixel Gaussian. Lower right: renzogram plot from the M33
    observation of \citealt{Putman09}}
  \label{fig:renzo_M33}
\end{figure*}

A column density plot is shown in Fig.~\ref{fig:ncol_M33} which shows
a more face-on view than M31 (see Fig.~\ref{fig:ncol_M31}). Renzogram
plots, overlaid on a grey scale plot of summed intensity, are shown in
Fig.~\ref{fig:synth_blur1_M33} and Fig.~\ref{fig:synth_blur6_M33} for
synthetic data blurred with a 1 pixel and 6 pixel Gaussian
respectively (1 pixel corresponds to 0.15 arcmin and 6 pixels
corresponds to 0.92 arcmin at a distance of 730~kpc).  The minimum
velocity is $-342$~km/s and contours increase in steps of 36~km/s to a
maximum of $-18$~km/s.
A corresponding contour plot from the observational data of
\cite{Putman09} is shown in Fig.~\ref{fig:obs_M33}. As M33 has a
smaller rotation velocity than our model galaxy (100 km/s compared to
220 km/s) the velocities which are contoured in Fig~\ref{fig:obs_M33}
have been scaled by a factor of 100/220 compared to the velocities
contoured in the synthetic data plots. For both the synthetic data and
the observed data the contour interval is 16 per cent of the rotation
velocity.

We again get a good qualitative comparison between the synthetic and
real observations, taking into account that we do not model the inner
5~kpc of the galaxy. The general behaviour of the contours is as
expected i.e. the main disc of the galaxy is traced out as velocity
increases and there are perturbations in the velocity contours due to
non-circular velocities.  Figure~\ref{fig:synth_blur1_M33} clearly
shows perturbations in the contours associated with the spiral arms
whereas lower resolution data (Fig.~\ref{fig:synth_blur6_M33}) show
much less detail but exhibit a broadening of the contours in the
region of the spiral arms. Our simulated galaxy is a grand design
spiral which shows non-circular motions due to the presence of spiral
shocks. In contrast M33 is a flocculent spiral with a relatively weak
spiral perturbation. The structure in M33 is instead largely due to
gravitational instabilities and supernovae feedback which are not
included in the model galaxy. Moreover even in the absence of spiral
shocks there are likely to be non-circular motions due to other
mechanisms (e.g. tidal interactions and warps). The simulated data
show what would be observed in an idealised case where a strong spiral
perturbation dominates the structure of the galaxy.

\subsection{Comparison to optically thin limit}

In the optically thin limit the absorption term in the radiative
transfer equation  can be neglected i.e. 
\begin{equation}
\frac{dI}{ds} = -\kappa I + \epsilon \approx \epsilon
\end{equation}
and the intensity in an image pixel can be found by simply integrating
the emissivity along the required line of sight. In this case the
intensity is proportional to the column density. As a validation test
of the method the ratio of intensity from the radiative transfer code
to the intensity derived from the optically thin approximation was
calculated for each pixel in the data cube. The ratio of these
intensities should be unity when the optical depth is small, and the
optically thin approximation holds, and should decrease below unity as
the optical depth increases.

\begin{figure*}
  \centering
  \subfigure[]{\includegraphics[scale=0.3]{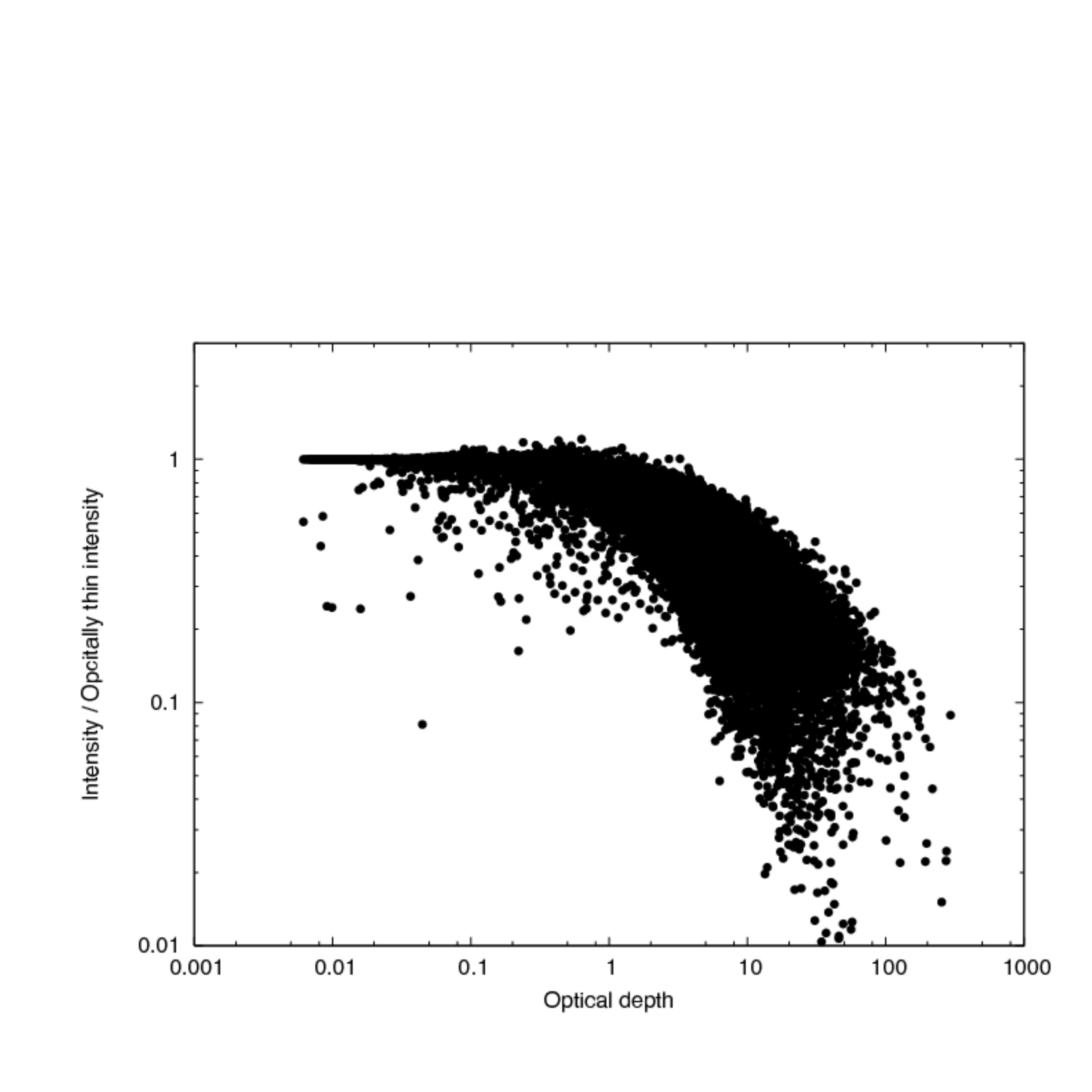}\label{fig:m31_ratio_250bins}}
  \subfigure[]{\includegraphics[scale=0.3]{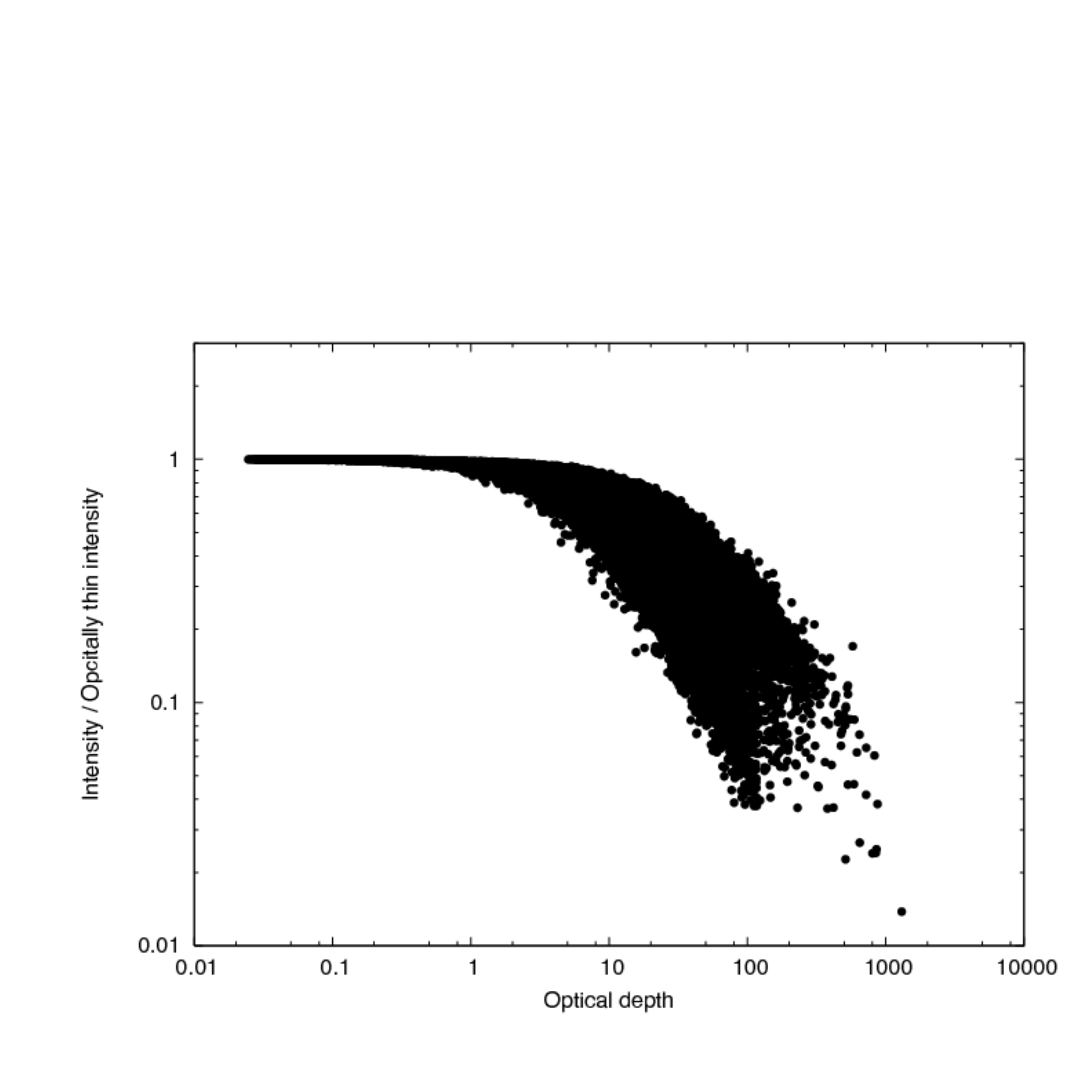}\label{fig:m31_ratio_1000bins}}
  \caption{Ratio of intensity from radiative transfer calculation to
    intensity from the optically thin approximation, plotted against
    optical depth (for the simulated M31). Left: from a calculation
    using 250 velocity bins. Right from a calculation using 1000
    velocity bins.}
  \label{fig:m31_ot_comp}
\end{figure*}

\begin{figure}
  \centering
  \includegraphics[scale=0.3]{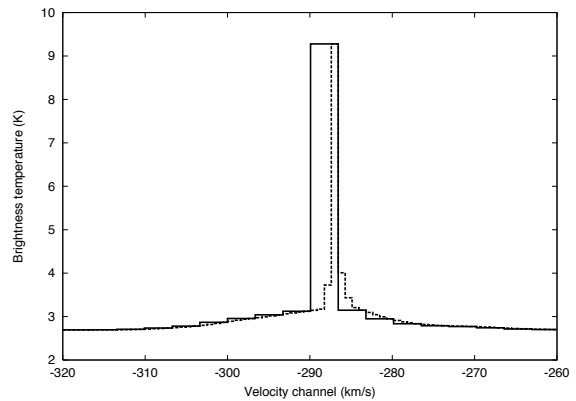}
  \caption{Line profiles from $x=-0.972$~kpc, $y=-0.3626$~kpc. The solid line is
    from a calculation using 250 velocity bins, the dashed line is
    from a calculation using 1000 velocity bins. The velocity
    integrated emission is overestimated in the 250 bin profile due
    to a lack of resolution in velocity space.}
  \label{fig:large_ratio_example}
\end{figure}

Figure~\ref{fig:m31_ot_comp} plots the ratio of intensity from the
radiative transfer calculation to intensity from the optically thin
approximation against optical depth for the simulated
M31. Figure~\ref{fig:m31_ratio_250bins} is from the results presented
in the previous section with 250 velocity bins over a range of
840~km/s. Pixels in which the column density is less than
$10^{20}~\rm{cm}^{-2}$ are not plotted in order to exclude pixels
which are not associated with the galaxy. The intensity ratio at low
optical depths is close to unity and there is a trend of decreasing
intensity, relative to the optically thin limit, as optical depth
increases, as expected. However there is a considerable amount of
scatter and there are some points with a ratio greater than unity, up
to a maximum value of 1.21. An example line profile from a pixel with
a ratio greater than unity is shown in
Fig.~\ref{fig:large_ratio_example}. This pixel is located at
$x=-0.972$~kpc, $y=-0.3626$~kpc and is not included in
Fig.~\ref{fig:m31_ot_comp}, as the column density is below the cutoff
threshold. However it is used here as a clear example of how the
velocity integrated emission in a pixel can be overestimated due to a
lack of resolution in velocity space.
Figure~\ref{fig:m31_ratio_1000bins} shows the same plot as
Fig,~\ref{fig:m31_ratio_250bins} but for the synthetic data cube with
1000 velocity bins. The points with ratios greater than unity are
absent and there is less scatter on the distribution. The velocity
resolution clearly has a significant impact on the results of this
validation test.

\begin{figure}
  \centering
  \includegraphics[scale=0.3]{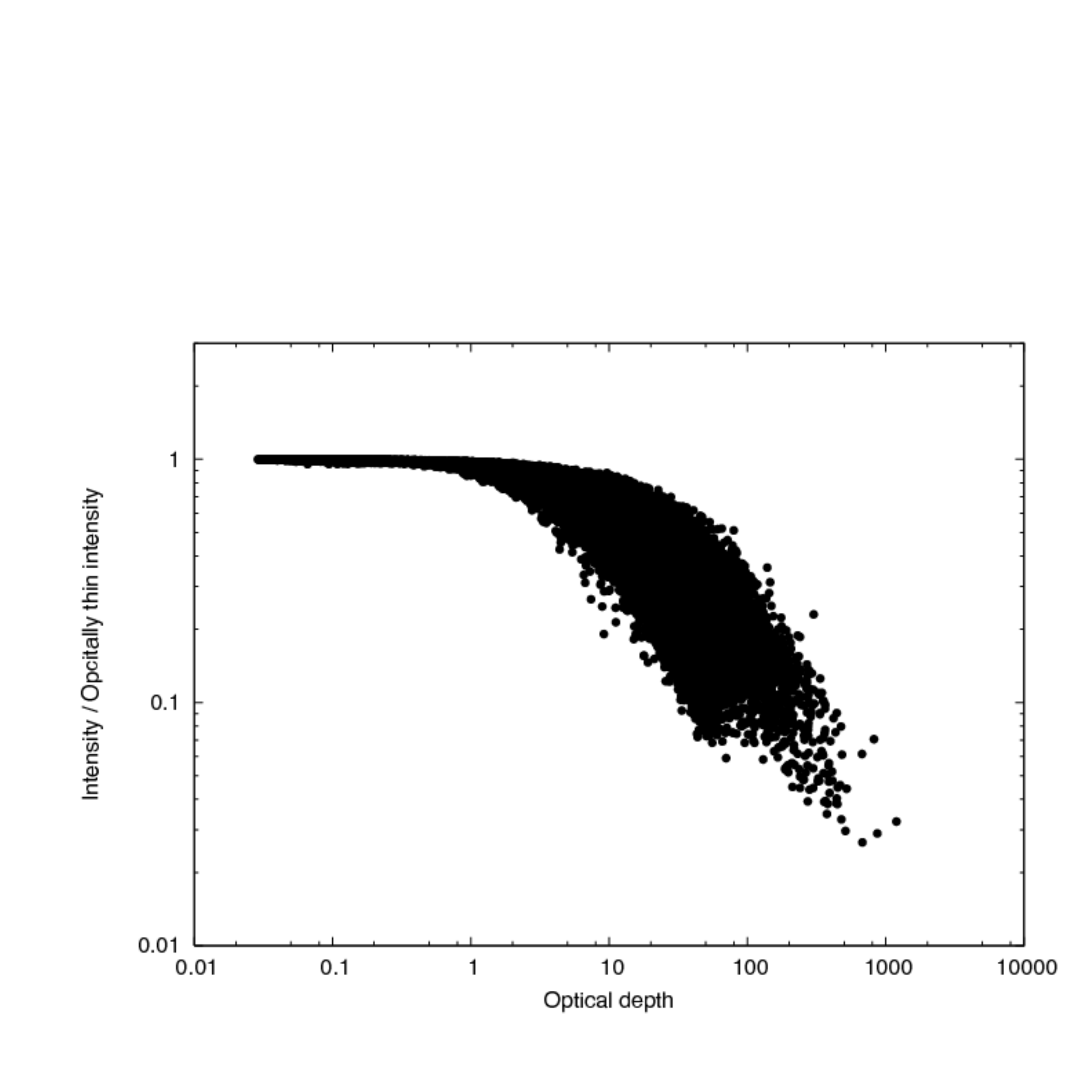}
  \caption{Ratio of intensity from radiative transfer calculation to
    intensity from the optically thin approximation plotted against
    optical depth for the simulated M33.}
  \label{fig:m33_ratio}
\end{figure}

A similar plot showing the ratio of the intensity from the radiative
transfer calculation to intensity from the optically thin
approximation against optical depth for the simulated M33 is shown in
Fig.~\ref{fig:m33_ratio}.
As before pixels with a column density less than
$10^{20}~\rm{cm}^{-2}$ are not plotted. The velocity resolution of the
synthetic M33 cube is high enough that the comparison with the
optically thin limit is good and does not show the scatter seen in
the M31 cube when only 250 velocity channels are output. 

\subsection{Velocity perturbations in renzograms}

We develop a simple analytical representation of the velocity
perturbations seen in the renzograms, in order to gain a quantitative
understanding of how changes in velocity within the galaxy relate to
perturbations in renzogram contours. The following analysis uses a
cylindrical polar co-ordinate system with its origin at the centre of
the galaxy.  The $r$ co-ordinate represents the distance from the
galactic centre in the midplane, the $z$ co-ordinate is the distance
out of the midplane and $\theta$ is the cylindrical polar angle. It is
assumed that all gas moves in the azimuthal direction only.

The line of sight velocity, towards the observer, of a parcel of gas is 
\begin{equation}
v_{\rm{los}} = v_{\rm{t}} \cos \theta \sin i + v_{\rm{sys}}
\end{equation}
where $v_{\rm{t}}$ is the magnitude of the tangential velocity
(velocity in the azimuthal direction) for a given parcel of gas, $i$
is the inclination of the galaxy and $v_{\rm{sys}}$ is the systemic
velocity of the galaxy. For an observation with the systemic velocity
subtracted this gas will be seen in velocity channel
 \begin{equation}
v_{\rm{ch}} = v_{\rm{t}} \cos \theta \sin i
\end{equation}
In the absence of any deviation from a constant circular velocity 
the renzogram contours have a constant value of $\theta$. 
However, the effect of the spiral shock means that in a given velocity channel
we expect to see both gas moving at the circular velocity of the galaxy
($v_t=v_c$) and other gas which has just passed through a spiral shock
(post-shock gas). This causes an azimuthal perturbation in the
renzogram contour. 
 
We assume that the effect of passing through the shock is to reduce
the speed of this gas to $v_t=f v_c$ ($0<f<1$) without changing its
direction. Gas moving at the circular velocity will be seen at a
position
\begin{equation}
\cos \theta_{c} = \frac{v_{\rm{ch}}}{v_c \sin i}
\label{eqn:theta_c}
\end{equation}
whereas post-shock gas is seen at a position
\begin{equation}
\cos \theta_{ps} = \frac{v_{\rm{ch}}}{f v_c \sin i}
\label{eqn:theta_ps}
\end{equation}
The linear separation in the plane of the disc ($d$) of circular and
post-shock gas is given by 
\begin{equation}
d = 2r \sin \left( \frac{\theta_c - \theta_{ps}}{2} \right)
\end{equation}
which can be re-written, using trigonometric identities and
substituting from eqn.~\ref{eqn:theta_c} and eqn.~\ref{eqn:theta_ps}, as 
\begin{equation}
d = 2r \left[ \left(  1-x^2 \right)^{1/2} \frac{x}{f} - x \left( 1 - \left(\frac{x}{f}\right)^2
  \right)^{1/2} \right]
\end{equation}
where 
\begin{equation}
x = \frac{v_{\rm{ch}}}{v_{\rm{c}} \sin i}
\label{eqn:x_param}
\end{equation}
When viewed on the plane of the sky this distance is reduced, due
to projection effects, by a factor of 
\begin{equation}
\left( \sin^2 \theta + \cos^2 i \cos^2 \theta \right)^{1/2} = 
\left( 1 + \left( \cos^2 i  -1 \right) x^2 \right)^{1/2}
\end{equation} 
Hence the angular size of the perturbation in the velocity contour ($\Delta \alpha$) is 
\begin{eqnarray}
\Delta \alpha = \frac{2r}{D_{\rm{gal}}} \left[ \left(  1-x^2 \right)^{1/2} \frac{x}{f} - x \left( 1 - \left(\frac{x}{f}\right)^2
  \right)^{1/2} \right] \nonumber \\
\times \left( 1 + \left( \cos^2 i  -1 \right) x^2 \right)^{1/2}
\label{eqn:angsize}
\end{eqnarray}
where $r$ is the distance of the perturbation from the galactic
centre and $D_{\rm{gal}}$ is the distance to the galaxy.

At $x=0$ (i.e. the channel corresponding to the systemic velocity)
there is no velocity perturbation according to
eqn.~\ref{eqn:angsize}. In this channel gas which is moving in
circular motion has no line of sight component. As we have assumed 
that the velocity perturbation is simply a reduction of the circular
motion there is no mechanism for changing the line of sight
component. In the renzograms derived from the synthetic observations
there are perturbations seen in this velocity channel 
indicating that the velocity perturbation in our model galaxy is not
simply a reduction of the circular motion but has another component
which affects the line of sight velocity.

\begin{figure*}
  \centering
  \subfigure[]{\includegraphics[scale=0.3]{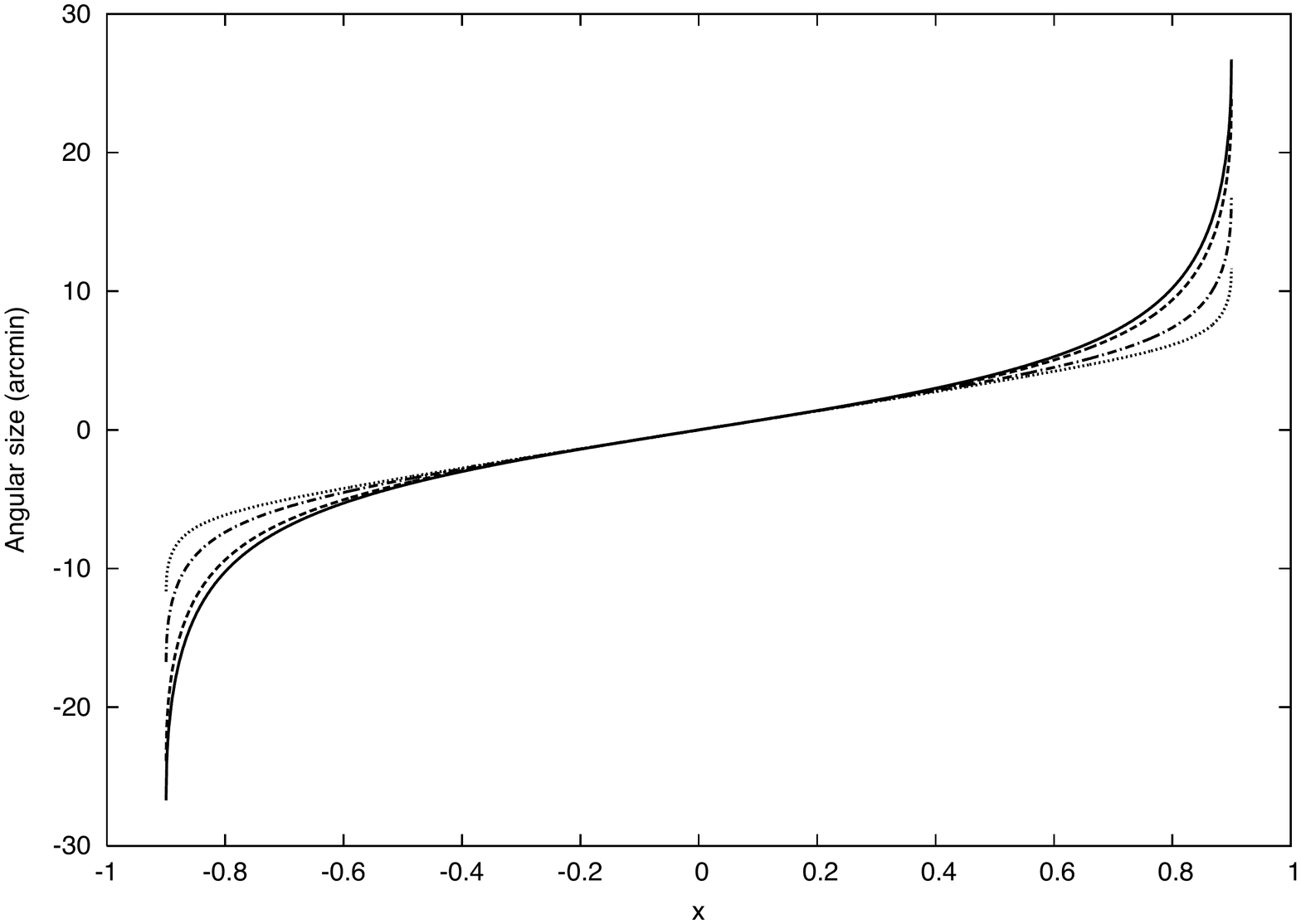}\label{fig:angsize}}
  \subfigure[]{\includegraphics[scale=0.3]{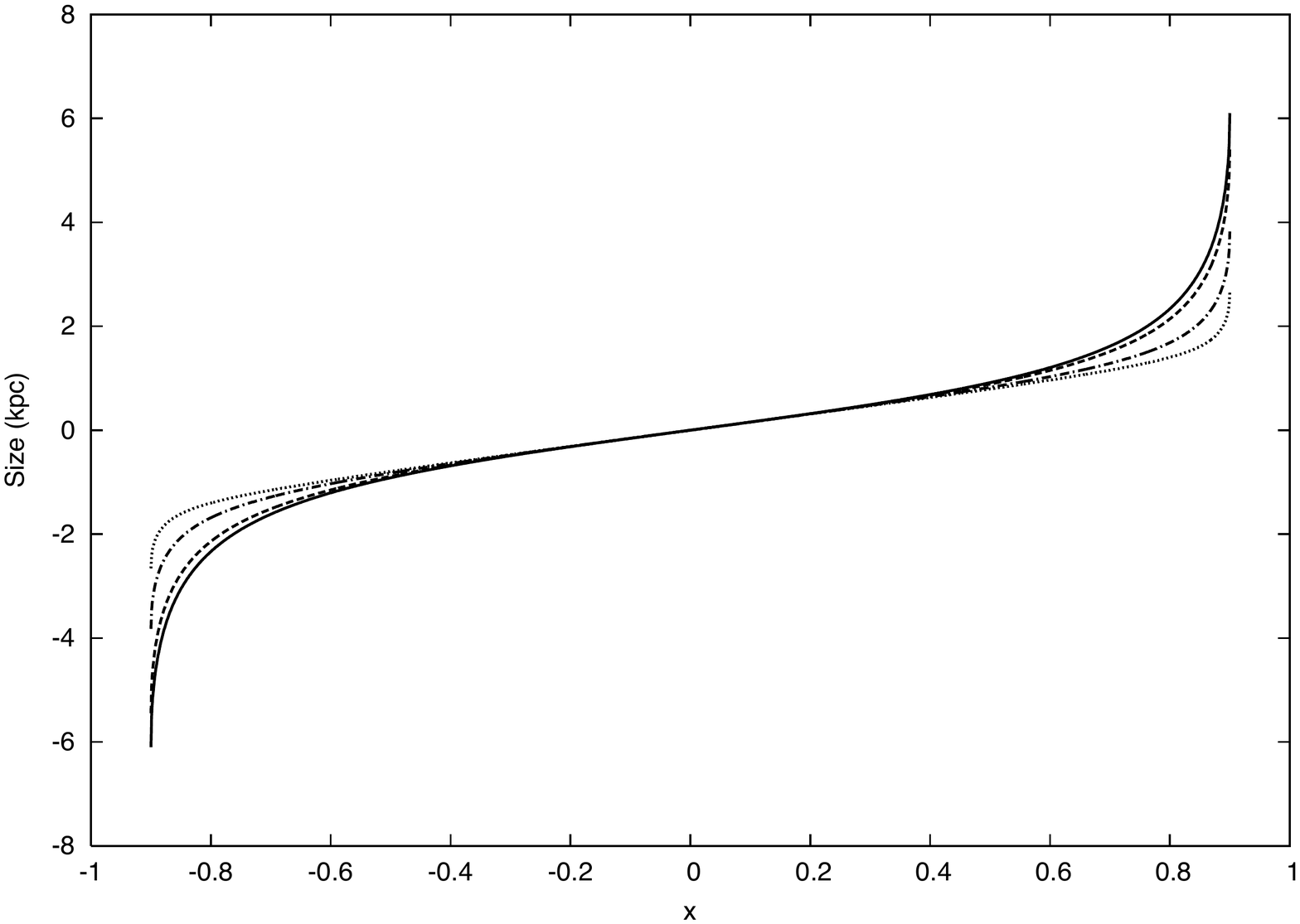}\label{fig:linsize}}
  \caption{Left: Angular size (in arcmin) of velocity perturbations, according to
    eqn.~\ref{eqn:angsize}, for a galaxy at 785~kpc with a shock at
    7~kpc from the galactic centre. The shock strength parameter is
    $f=0.9$. Four inclinations are shown: 0~degrees (solid line),
    30~degrees (dashed line), 60~degrees (dot-dashed line) and
    90~degrees  (dotted line). Right: as for previous figure but with
    sizes in kpc. The parameter $x$ (see eqn~\ref{eqn:x_param}) is the velocity channel scaled by
    the circular velocity projected onto the line of sight ($v_c \sin i$).}
\end{figure*}

The angular size of the velocity perturbations seen in the renzograms
(as predicted by eqn~\ref{eqn:angsize}) is
plotted against $x$ in Fig.~\ref{fig:angsize} for four inclination
angles; 0~degrees (solid line), 30~degrees (dashed line), 60~degrees
(dot-dashed line) and 90~degrees (dotted line). This plot assumes the
galaxy is at a distance of 785~kpc (i.e the distance to
M31) with a shock 7~kpc from the galactic centre (representative of
our model galaxy). The shock is assumed to reduce the circular
velocity by a factor $f=0.9$ (i.e. a perturbation of 20~km/s on a
200~km/s circular velocity). A similar plot showing the size of the
velocity features in kpc is shown in Fig.~\ref{fig:linsize}. The predicted
size of the features is in agreement with the features seen in the
renzograms of the synthetic observations (i.e. a typical size of
approximately a kpc). 
In all cases the size of the velocity perturbation tends towards zero
as $x$ tends towards zero, due to the assumption that the perturbation
only affects the circular velocity (as discussed above). As $x$
increases towards $f$ the size of the feature increases because the
circular velocity has a larger component in the line of sight. The
renzograms for our synthetic observations also show larger velocity
kinks in channels further from the systemic velocity. There is a
change in the sign of $\Delta \alpha$ when $x$ becomes negative
(eqn~\ref{eqn:angsize} is an odd function of $x$) due to the
assumption that the shock always reduces the circular velocity. This
results in post-shock gas always moving towards the velocity channel
of the systemic velocity (or zero if the systemic velocity has been
subtracted). The size of the velocity kinks increases as the
inclination angle increases, due to projection effects, however in
practice it will be difficult to see these features for nearly face-on
cases because very high velocity resolution would be required (for
nearly face-on cases the range of $x$ corresponds to a small velocity
range).

\section{Conclusions and future work}

We have presented a method for generating synthetic spectral cubes of
the 21~cm hydrogen line from SPH simulations of spiral galaxies. The
method successfully maps the density, temperature and velocity from
the SPH particles onto an AMR grid, while preserving important
structures (e.g. spiral arms and spurs) and accurately representing
the total mass. Synthetic data cubes are generated using a ray tracing
method. The synthetic data show good agreement with observations of
M31 and M33 whereby increasing velocity channels trace out the main
disc of the galaxy. Velocity contours of the synthetic data show
perturbations due to non-circular motions, similar to those already
observed in M81 \citep{Adler96,Visser1980,rots_1975}, which are not
seen in the observations of M31 and M33. 

Our model galaxy is a grand design spiral galaxy and shows
    velocity structure associated with the spiral perturbation.  The
    method of generating synthetic observations can also be applied to
    simulations in which velocity structure is generated by other
    mechanisms. Internal mechanisms, such as self-gravity and stellar
    feedback, can generate velocity structure, and indeed are dominant
    in flocculent spiral galaxies.  External influences can also
    affect the {H\sc{i}} morphology of a spiral galaxy
    e.g. interactions with a companion galaxy, high velocity clouds or
    the intracluster medium.  As the majority of galaxies reside in
    groups or clusters, it is likely that external environmental
    effects will influence the {H\sc{i}} structure of most real
    galaxies; our model galaxy is perfectly isolated and our results
    show idealised behaviour in the absence of external influences.  

A large parameter space could potentially be studied, involving
    galaxies with different internally and externally generated
    velocity structure. Some of these aspects can already be modelled
    (e.g. a galaxy undergoing an interaction, \cite{Dobbs09}) and some
    require further model development (e.g inclusion of stellar
    feedback, which is ongoing). A large number of models must be be
    run in order to generate the SPH input for the radiative transfer
    calculation. Once such a library of simulations was available, one
    could compare these with real observations in order to understand
    the velocity structure of spiral galaxies. However given the
    potentially complex nature of the velocity structures generated
    there would need to be a reliable way of matching an observed
    galaxy with its counterpart in the synthetic observation
    library. In our present models, we can relate the strength of the spiral
    shock, and the inclination of the galaxy, to the size of the
    perturbations in the renzograms. When we include
    additional physics, e.g. self gravity and feedback, we can
    investigate whether these perturbations are still observable (for
    a given shock strength) or whether they are overwhelmed by other
    motions in the gas.

Our method has also recently been applied to an observer placed inside
the galaxy \citep{douglas_2010} to generate a synthetic galactic plane
survey.  H\,{\sc{i}} self absorption features and the conversion of
cold H\,{\sc{i}} to molecular clouds, as seen in a real galactic plane
survey (e.g. \cite{Taylor03,Stil06}), are seen in the synthetic
data. A powerful future application of using a simulation for
generating a survey is that given a series of time frames we will be
able to trace the evolution of molecular clouds, and related observed
features, to the physical properties of the material in the cloud. The
method can also be extended to generate synthetic observations of
other tracer species, such as CO, or dust, for simulated surveys or
simulated external galaxies.

\section*{Acknowledgments}
We would like to thank Tyler Foster for supplying the M31 data used in
this paper. We would like to thank Laurent Chemin for providing the
M31 observed line profile data.  Calculations presented here were
performed using the University of Exeter Supercomputer. The research
leading to these results has received funding from the European
Community's Seventh Framework Programme under grant agreement n$^o$
PIIF-GA-2008-221289. C.~L.~D.'s research at Exeter was conducted as
part of the award ``'The formation of stars and planets: Radiation
hydrodynamical and magnetohydrodynamical simulations,'' made under the
European Heads of Research Councils and European Science Foundation
EURYI (European Young Investigator) Awards scheme and supported by
funds from the Participating Organisations of EURYI and the EC Sixth
Framework Programme.

\bibliographystyle{mn2e}
\bibliography{acreman_2010} 

\end{document}